\documentclass[aps,pre,epsf,superscriptaddress,amsmath,amssymb,amsfonts,twocolumn,showpacs]{revtex4-1}

\usepackage{graphicx}
\usepackage{epsfig}
\usepackage{dcolumn}
\usepackage{bm}
\usepackage{braket}
\usepackage{amsmath}
 \usepackage{color}
 \newcommand{\abs}[1]{\left| #1 \right|}
\begin{document}
\title{Dark-Bright Soliton  Dynamics \\ Beyond the Mean-Field
  Approximation}

\author{G.C. Katsimiga}
\affiliation{Zentrum f\"{u}r Optische Quantentechnologien,
Universit\"{a}t Hamburg, Luruper Chaussee 149, 22761 Hamburg,
Germany}  
\author{G.M. Koutentakis}
\affiliation{Zentrum f\"{u}r Optische Quantentechnologien,
Universit\"{a}t Hamburg, Luruper Chaussee 149, 22761 Hamburg,
Germany}\affiliation{The Hamburg Centre for Ultrafast Imaging,
Universit\"{a}t Hamburg, Luruper Chaussee 149, 22761 Hamburg,
Germany}
\author{S.I. Mistakidis}
\affiliation{Zentrum f\"{u}r Optische Quantentechnologien,
Universit\"{a}t Hamburg, Luruper Chaussee 149, 22761 Hamburg,
Germany}
\author{P. G. Kevrekidis}
\affiliation{Department of Mathematics and Statistics, University
of Massachusetts Amherst, Amherst, MA 01003-4515, USA }
\author{P. Schmelcher}
\affiliation{Zentrum f\"{u}r Optische Quantentechnologien,
Universit\"{a}t Hamburg, Luruper Chaussee 149, 22761 Hamburg,
Germany} \affiliation{The Hamburg Centre for Ultrafast Imaging,
Universit\"{a}t Hamburg, Luruper Chaussee 149, 22761 Hamburg,
Germany}

\date{\today}

\begin{abstract}
  The dynamics of dark-bright solitons
  beyond the mean-field approximation is investigated. We first examine the case of a single dark-bright
  soliton and its oscillations within a parabolic trap. Subsequently,
  we move to the setting of collisions, comparing the mean-field
  approximation to that involving multiple orbitals in both the dark
  and the bright component. 
  Fragmentation is present
  and significantly affects the dynamics, especially in the
  case of slower solitons and in that of lower atom numbers.
  It is shown that  
  the presence of fragmentation 
  allows for bipartite entanglement 
  between the distinguishable species. 
  Most importantly the interplay between fragmentation and entanglement leads to 
  the splitting of each of the parent mean-field dark-bright solitons, placed off-center within the parabolic trap, 
  into a fast and a slow 
  daughter solitary wave.   
  The latter process is in direct contrast to the predictions of the mean-field approximation. 
  A variety of excitations including dark-bright
  solitons in multiple (concurrently
  populated) orbitals is observed. Dark-antidark states and domain-wall-bright 
  soliton complexes can also be observed to arise spontaneously
  in the  beyond mean-field dynamics.
\end{abstract}

%\pacs{Pacs numbers} 
\maketitle

\section{introduction}

Over the last two decades, atomic Bose-Einstein condensates (BECs)
have offered an ideal
testbed for the exploration of coherent structures
relevant to nonlinear wave systems~\cite{stringari,siambook}.
Most of these entities have not only been theoretically
predicted but also in many case examples experimentally verified \cite{experiment,experiment1,experiment2}. 
Some of the most notable ones are the
bright~\cite{expb1,expb2,expb3},
dark~\cite{djf} and gap~\cite{gap} matter-wave solitons.
Higher dimensional analogues of these structures have also been
examined, such as  vortices~\cite{fetter1,fetter2},
solitonic vortices and vortex rings~\cite{komineas_rev}.

Although the main emphasis of study has been the case of single
component systems, in recent years, the study of multi-component
variants has also gained a significant traction~\cite{siambook}.
There, new fundamental waveforms may arise such as the symbiotic
dark-bright (DB) soliton in self-defocusing media (where a bright
wave would not exist, yet it survives being confined by
the dark component)~\cite{revip}.
Interestingly, the experimental study of such states was initiated
considerably earlier
in nonlinear optics, per the observation of DB
solitary wave structures and their molecules~\cite{seg1,seg2}. 
More recently, atomic BECs enabled a wide variety
of relevant studies initially motivated by the theoretical
study of~\cite{buschanglin}.
In particular, the experimental realization of DBs~\cite{hamburg}
was followed by a series of experiments exploring
the dynamics and properties of these states
including in-trap oscillations of the solitons,
their spontaneous generation (e.g.
via counterflow experiments) and their interactions both with other
DBs and with external potential barriers~\cite{pe1,pe2,pe3,pe4,pe5,azu}.

A natural question that has emerged in connection with
solitary waves concerns the fate
of these types of excitations in the presence of quantum
fluctuations~\cite{Drummond1,Drummond2,Drummond3,Drummond4}. 
In atomic BECs, this has been considered for bright
solitons~\cite{castin1,castinbr,carrbrand,sacha1,cederbaum,bettina,delande,billam,Ilg,weiss1}, 
and dark solitons~\cite{sachadark3,sachadark2,sachadark1,sacha,sven},
in trapped settings, 
as well as in the presence of
optical lattices~\cite{mishmash,barbiero}.  
A number
of these works, including~\cite{martin} among others, has also
explored the role of quantum fluctuations and higher
orbitals in the presence of solitonic interactions.
Many of the resulting findings suggest that slower
solitons and smaller atom numbers result in significant
deviations from the mean-field, Gross-Pitaevskii (GP)
limit.

However, to the best of our knowledge, such studies have
not been performed in a systematic fashion in multi-component
settings and for associated solitary wave structures.
In that light, herein we explore the case of DB
solitons and their dynamics, as well as collisions both
at and beyond the mean-field limit.
To incorporate the quantum fluctuations stemming 
from the correlations in the DB soliton dynamics,
we employ the Multi-Layer Multi-Configuration 
Time-Dependent Hartree Method for bosons (ML-MCTDHB)~\cite{Sven,Lushuai}
designed for simulating the quantum dynamics
of bosonic mixtures.
We consider both
the oscillation of a single DB solitary wave in a 
trap, as well as the interaction of two symmetric solitary waves inside a parabolic
trap. We compare and contrast the findings of the mean-field
case (where a single orbital is effectively used in the dark-
and bright- components) with cases where multiple orbitals  
are used. 
In all cases it is found that the initial mean-field DB solitons split into  
daughter DB solitary waves, in contrast to the well-known mean-field predictions.   
The robustness of the presented results is ensured by exploring settings 
that involve a higher number of orbitals thereby supporting the validity of our approximation and of the observed beyond mean-field excitations (see also Appendix B).  
Within the employed multi-orbital 
approximation bipartite entanglement (see \cite{Horodecki,Horodecki1} and references therein) between the distinguishable species 
resulting from the spontaneous fragmentation of the DBs is generally present.
More importantly it is the interplay between fragmentation and the resulting
entanglement which gives rise to the observed dynamical structures. 

The dynamics of a single DB soliton being initialized off-center in the parabolic trap, is first analyzed.
It is shown that from the early stages of the dynamics the initial DB solitary wave becomes both fragmented 
and entangled, and consequently splits into two DB solitonic fragments
signalling its decay. This is in direct contrast with the pure mean-field description
where the DB soliton remains intact for large propagation times.
After the splitting process novel structures of multi-orbital nature emerge, 
such as dark-antidark states~\cite{ionut,Kevrekidis1},
with the latter corresponding to density humps on top of the BEC background,
and domain-walls (DW)~\cite{stringari,siambook}. The corresponding decay time  
increases for larger initial velocities thus approaching the mean-field limit. Furthermore the lifetime of the DB entity    
decreases for larger particle number imbalance (i.e. varying the number of atoms 
in the dark soliton component upon keeping fixed the number of atoms in the respective bright counterpart).  
As a next step the collisional dynamics between two DBs is 
investigated.
We find that in this case too, both fragmentation and entanglement are evident from the beginning of the dynamics.  
Perhaps more intriguingly, a variety of unprecedented
(and some unique to the multi-component setting) states are identified,
even if in a transient form during our dynamical evolutions.
These include among others DB solitons in {\it higher}
orbitals alone, DB solitons involving both the lower and
higher orbitals, DWs,
as well as dark-antidark type structures
between the orbitals of the same
(dark) component.
It is important to note that in the cases reported
significant fractions of populations arise in higher orbitals
even among the 500 or 1000 atoms present in the dark component.

Our presentation proceeds as follows. Section II 
provides an overview of the setup and background, 
in which we discuss
the preparation of the DB solitons and related states.  
Next, we present and discuss our numerical results for the single DB soliton 
dynamics (section III) and the DB soliton collisions (section IV). 
Finally, in section V, we provide our 
conclusions.
Appendix A focuses 
on the initial state preparation, while Appendix B 
is dedicated to a discussion of our computational approach and 
the convergence of our simulations.

\section{Setup and background knowledge}

Dark-bright solitons are non-linear excitations observed in one-dimensional (1D) population-imbalanced binary BECs.
Such excitations are characterized by a density depletion for the majority component (that we will hereafter dub species $D$) 
of the BEC, accompanied by a density accumulation 
for the minority component (hereafter referred to as
species $B$). 
In the mean-field approximation the prototypical 1D model where such states can be found
to arise \cite{pe3}, is a system of coupled GP equations, 
i.e. a vector variant of the nonlinear Schr{\"o}dinger equation \cite{Pethick,Gross,Dalfovo} with cubic nonlinearity. 
Each DB soliton is characterized by its position $x_j^{0}$,
its inverse width $d_j$, 
and by the so-called soliton's phase angle $a_j$. The latter is associated with the soliton's
velocity $u_j/c = \sin a_j$, with $c=\sqrt{gn/m}$, being the speed of sound. 
Here, $g$ denotes the interparticle interaction, $n$ refers to the local particle density, and $m$ 
is the particle mass. 

In the following we assume $N_S$ DB solitons (with $N_S=1$ and $N_S=2$ corresponding to a single and two DB solitons 
respectively) 
embedded in a background
density, $\tilde\phi_0(x)$. The wavefunction of the majority (dark) component~\cite{gasses} reads 
\begin{equation}
\begin{split}
\tilde \phi^D(x;t)\equiv \tilde\phi_0(x) \prod_{j=1}^{N_S} \Big[ &\cos a_j~ \tanh \left[ d_j \left( x-x_{j}(t) \right)\right] \\
&+ i \sin a_j \Big], \label{Eq:1}
\end{split}
\end{equation} 
with $x_{j}(t)=x_{j}^0-u_j t$.
The corresponding wavefunction for the minority (bright) component and upon considering in-phase bright counterparts
reads 
\begin{equation}
\tilde \phi^B(x;t)\equiv\sum_{j=1}^{N_S}~  \eta_j \text{sech }\left[ d_j \left( x-x_{j}(t) \right)\right]~ e^{i \left( d_j 
\tan a_j x-\theta_j(t)\right)}, \label{Eq:2}
\end{equation}
where $\theta_j(t)=1/2 \left(d^2_j-d^2_j\tan^2 a_j\right)t+\left(\mu'-\mu\right)t$. Note that, $\mu'$
is the chemical potential of the bright component, being fixed so as to correspond to a total number of atoms for the bright 
counterpart $N_B=5$, that we also keep fixed throughout this work. Furthermore, 
$\eta_j$ refers to the amplitude of the bright soliton. In the absence of a confining potential
and for $N_S=1$, this amplitude is related to  
the amplitude of the dark soliton, $\sqrt{\mu} \cos a_j$, where $\mu$
is the background chemical potential, and the inverse width $d_j$ by  
\begin{equation}
\eta_j= \sqrt{\mu ~\cos^2 a_j - d_j^2 }. \label{Eq:3}
\end{equation} 
The above expressions namely Eqs.~(\ref{Eq:1})-(\ref{Eq:2}), represent an {\it approximate} initial profile of the
mean-field (i.e. single orbital) setting. The approximate nature of the profile
stems from the effective multiplication with the equilibrium background
$\tilde\phi_0(x)$ (at least in the case where $\tilde\phi_0(x)$ is not a constant),
which in the Thomas-Fermi limit reads: $\tilde\phi_0(x)=\left(1/\sqrt{g}\right)\sqrt{\mu-V(x)}$.
Here, $V(x)$ refers to the external trapping potential. We note also here, that in the case of collisions to be presented below,
  the two DB solitons ($N_S=2$) should be far enough apart to be individual entities, i.e., their width (proportional to $1/d_j$) should be much smaller
  than their relative separation $|x_i-x_j|$. 

In addition to the above approximation, the realm of the mean-field
itself leads to the 
following two assumptions (irrespectively of configuration): 
(a) the two species of the binary BEC are uncorrelated 
and (b) the constituting particles of each component are uncorrelated.
Therefore, the total many-body wavefunction within the mean-field approximation 
is expressed in terms of the mean-field wavefunctions 
\begin{equation}
\begin{split}
\Psi_{MF} (\vec x^D,\vec x^B;t) &= \Psi^D_{MF} (\vec x^D;t) \Psi^B_{MF} (\vec x^B;t)\\
&= \prod_{i=1}^{N_D} \frac{\phi^D(x^D_i;t)}{\sqrt{N_D}} \prod_{i=1}^{N_B} \frac{\phi^B(x_i;t)}{{\sqrt{N_B}}}, \label{Eq:4}
\end{split}
\end{equation} 
where $N_D$ ($N_B$) refers to the number of atoms for the $D$ ($B$) species, while 
$\vec x^D=\left( x^D_1, \dots, x^D_{N_D} \right)$ and $\vec x^B=\left( x^B_1, \dots, x^B_{N_B} \right)$ label the (spatial) coordinates of the atoms. 
$\phi^{D(B)}(x^{D(B)}_i;t)$ denotes the time-evolved wavefunction for the species $D$ ($B$) respectively within the mean-field approximation.  
The equations of motion for the mean-field ansatz [see Eq.~(\ref{Eq:4})] yield the well-studied  
GP equation. 

The binary BEC consisting of species $D$, $B$ (with Hilbert spaces $\mathcal{H}^D$, $\mathcal{H}^B$ respectively) is a bipartite composite 
system with the corresponding Hilbert space $\mathcal{H}^{DB}=\mathcal{H}^D\otimes\mathcal{H}^B$.   
To examine the system of $N_S$ solitons beyond the mean-field approximation, we incorporate 
the inter- and intra-species correlations by introducing $M$ distinct species functions for each component
of the binary BEC. Then the many-body wavefunction $\Psi_{MB}$ can be expressed according to the Schmidt decomposition 
\begin{equation}
\Psi_{MB}(\vec x^D,\vec x^B;t) = \sum_{k=1}^M \sqrt{ \lambda_k(t) }~ \Psi^D_k (\vec x^D;t) \Psi^B_k (\vec x^B;t), \label{Eq:5}
\end{equation}
where the coefficient $\lambda_k(t)$ is referred to as the natural occupations of the species function $k$. We remark that 
due to $M<\min(\dim(\mathcal{H}^D),\dim(\mathcal{H}^B))$ the above expansion (see Eq. (\ref{Eq:5})) corresponds to a truncated 
Schmidt decomposition of rank $M$ \cite{Horodecki}.   
It is also important to mention that a state of the bipartite system (see Eq. (\ref{Eq:5})) cannot be expressed as a direct 
product of two states from the two subsystem Hilbert spaces $\mathcal{H}^D$, $\mathcal{H}^B$ if at least two coefficients 
$\lambda_k(t)$ are nonzero. In the latter case the system is referred to as entangled \cite{note_ent,Roncaglia}.  
Note that a particular particle configuration of species $D$ (represented by $\Psi_k(\vec x^D;t)$) is accompanied by a 
particular particle configuration of species $B$ (denoted by $\Psi(\vec x^B;t)$) and vice-versa \cite{Johannes}. 
A corresponding measurement of one of the species states e.g. $\Psi_{k'}^{D}$ collapses the wavefunction of the other species 
to $\Psi_{k'}^{B}$ thus manifesting 
the bipartite entanglement \cite{Peres,Lewenstein1}.   
In this way, the above ansatz constitutes an expansion for the many-body wavefunction $\Psi_{MB}$ in terms of inter-species 
modes of entanglement. 
In the following we shall refer to $\sqrt{\lambda_k(t)} \Psi^D_k (\vec x^D;t)\Psi^B_k (\vec x^B;t)$ as the $k$-th mode of 
entanglement. 
We remark that in the case of $\lambda_1(t_0)=\lambda_2(t_0)=1/2$ the many-body state of the system 
$\Psi_{MB} \left(\vec x^D,\vec x^B;t_0\right)$   
at time $t_0$ is referred to as a Schr{\"o}dinger cat state. 
The particle correlations are included by constructing each of
the species functions $\Psi^{\sigma}_k \left(\vec x^{\sigma};t\right)$ using the permanents of $m^{\sigma}$ distinct time-
dependent single particle functions (SPFs, $\varphi_1,\dots,\varphi_{m^{\sigma}}$) 
\begin{equation}
\begin{split}
&\Psi_k^{\sigma}(\vec x^{\sigma};t) = \sum_{\substack{n_1,\dots,n_{m^{\sigma}} \\ \sum n_i=N}} c_{k,(n_1,
\dots,n_{m^{\sigma}})}(t)\times \\ &\sum_{i=1}^{N_{\sigma}!} \mathcal{P}_i
 \left[ \prod_{j=1}^{n_1} \varphi_1(x_j;t) \cdots \prod_{j=1}^{n_{m^{\sigma}}} \varphi_{m^{\sigma}}(x_j;t) \right], 
 \label{Eq:6}
 \end{split}
\end{equation} 
where $\mathcal{P}$ is the permutation operator exchanging the particle configuration within the SPFs, 
$c_{k,\left(n_1,\dots,n_{m^{\sigma}}\right)}(t)$ 
denote the time-dependent
expansion coefficients of a particular permanent. 
$N_{\sigma}$ refers to the particle number within species ${\sigma}$ 
and $n_i(t)$ is the occupation number of the SPF $\varphi_i(\vec{x};t)$. 
Therefore the $N_{\sigma}$-body Hilbert space of species $\sigma$ is approximated by $\mathcal{H}^{\sigma} \sim 
S\left(\mathcal{H}_{\varphi^{\sigma}(t)}^{\otimes N_{\sigma}}\right)$, where 
$S$ is the symmetrization operator and $\mathcal{H}_{\varphi^{\sigma}(t)}$ the single particle Hilbert space spanned by the 
SPFs.  
Following a time-dependent variational principle, e.g. the Dirac Frenkel \cite{Frenkel,Dirac} or the McLachlan variational 
principle~\cite{McLachlan} 
for the generalized ansatz [see Eqs.~(\ref{Eq:5}), (\ref{Eq:6})]
yields the ML-MCTDHB \cite{Sven,Lushuai,Johannes,note1} equations of motion. 
This consists of a set of $M^2$ ordinary (linear) differential equations of motion for the coefficients $\lambda_k(t)$,   
coupled to a set of $M(\frac{(N_D+m^D-1)!}{N_D!(m^D-1)!}+\frac{(N_B+m^B-1)!}{N_B!(m^B-1)!})$     
non-linear integro-differential equations for the species functions $\Psi_k^{D(B)}(\vec{x}^{D(B)};t)$ and $m_D+m_B$ nonlinear 
integro-differential 
equations for the SPFs $\varphi_k(\vec{x};t)$. 
To the best of our knowledge analytical solutions of the many-body ansatz [Eqs. (\ref{Eq:5}), (\ref{Eq:6})] that contain DB 
solitons are not known while systematic numerical studies in this direction are still lacking. 
Here, we utilize the many-body approach that ML-MCTDHB provides and embed at $t=0$ the mean-field wavefunction [see 
Eq.~(\ref{Eq:4})] 
within the many-body ansatz [see Eqs.~(\ref{Eq:5}), (\ref{Eq:6})]. To achieve the latter, we consider $\lambda_1(0)=1$, 
$\lambda_{i \neq 1}(0)=0$ for the natural occupations of the species 
functions, $c_{1, n_1=N_{D}}(0)=1$, $c_{1, n_1 \neq N_D}(0)=0$ ($c_{1, n_1=N_{B}}(0)=1$, $c_{1, n_1 \neq N_B}(0)=0$) [see Eqs. 
(\ref{Eq:1}), (\ref{Eq:2})] for the expansion coefficients of 
species $D$ ($B$) and 
$\varphi^D_1(x;0)= \frac{1}{\sqrt{N_D}} \tilde \phi^D(x)$ ($\varphi^B_1(x;0)= \frac{1}{\sqrt{N_B}} \tilde \phi^B(x)$) for the 
SPFs of the majority (minority) component. 
The mean-field ground-state is used as the background density, $\tilde\phi_0(x)$.
Summarizing, we initialize the many-body quantum dynamics problem  
employing the mean-field initial state,  
aiming to examine how the 
single-orbital population will spontaneously give rise to 
higher orbital dynamics. 

For binary mixtures the one-body density can be expanded in different modes stemming from the species functions 
expansion [see Eq.~(\ref{Eq:5})]. For instance, the one-body density of the majority species 
$D$ reads 
\begin{equation}
\begin{split}
&\rho^{(1),D}(x,x';t)=\int d^{N_D-1}\bar x^D d^{N_B} x^B \times ~\\&\Psi^{*}_{MB}(x,\vec{\bar x}^D,\vec x^B;t) \Psi_{MB}(x',
\vec{\bar x}^D,\vec x^B;t)\\
&=\sum_{k=1}^M \lambda_k(t) \int d^{N_D-1}\bar x^D ~ \Psi^{*D}_k(x,\vec{\bar x}^D;t) \Psi^{D}_k(x',\vec{\bar x}^D;t)\\
&=\sum_{k=1}^M \lambda_k(t)~ \rho^{(1),D}_k(x,x';t),
\end{split} \label{Eq:8}
\end{equation} 
where $\rho^{(1),D}_i(x,x';t)$ refers to the one-body density matrix of the $i$-th species function
and $\bar x^D=\left(x^D_1,x^D_2,\ldots,x^D_{N_D-1}\right)$. The one-body density for the minority 
species $\rho^{(1),B}(x,x';t)$ can be defined in a similar manner. 
The natural orbitals, $\phi^{\sigma}_i(x;t)$, are defined as the eigenfunctions of the  
one-body density matrix $\rho^{(1),{\sigma}}(x,x')$. For our purposes we consider them to be normalized to their 
corresponding eigenvalues, $n^{\sigma}_i$ (natural occupations) as 
\begin{equation}
n^{\sigma}_i(t)= \int d x~ \left| \phi^{\sigma}_i(x;t) \right|^2. \label{Eq:9}
\end{equation}
In this way, we can ensure that when our many-body wavefunction $\Psi_{MB}(\vec x^D,\vec x^B;t)$ reduces to the mean-field 
case (i.e. $\Psi_{MB}(\vec x^D,\vec x^B;t) \to \Psi_{MF}(\vec x^D,\vec x^B;t)$)
the corresponding natural occupations obey $n_1^{\sigma}(t)=N^{\sigma}$, $n_{i\neq1}^{\sigma}(t)=0$.  
In the latter case the first natural orbital $\phi^{\sigma}_1(x^{\sigma};t)$ reduces to the mean-field wavefunction 
$\phi^{\sigma}(x^{\sigma};t)$. 
To examine the beyond-mean-field dynamics of a DB soliton
in a setting relevant to most of the recent experiments,
we consider a binary bosonic gas trapped in a 1D harmonic oscillator potential. 
The many-body Hamiltonian consisting of $N_D$, $N_B$ bosons with mass $m_D$, $m_B$ for the species $D$, $B$ respectively,  
reads 
\begin{equation}
\begin{split}
H= &\sum_{\sigma=D,B} \sum_{i=1}^{N_{\sigma}} \left[ -\frac{\hbar^2}{2 m_{\sigma}} \left( \frac{d}{d x^{\sigma}_i} \right)^2 %
+\frac{1}{2} m_{\sigma} \omega_{\sigma}^2  \left( x^{\sigma}_i \right)^2 \right]\\ 
&+\sum_{\sigma=D,B}g_{\sigma\sigma} \sum_{i<j} \delta( x^{\sigma}_i - x^{\sigma}_j)\\ 
%+ g_{BB} \sum_{i<j} \delta( x^B_i - x^B_j)\\
&+g_{DB} \sum_{i=1}^{N_D} \sum_{j=1}^{N_B} \delta( x^D_i - x^B_j).
\end{split}
\label{Eq:10}
\end{equation}
Within the ultracold $s$-wave scattering limit~\cite{Pethick}, we model both the inter- and intra-species with contact 
interaction by a delta function with respect to the 
relative coordinate of two bosons and the strength being denoted by $g_{DB}$, $g_{DD}$, $g_{BB}$ respectively. 
We also assume that the bosons of both species possess the same mass, i.e. $m_D$=$m_B$=$m$ and 
experience the same external potential, i.e. $\omega_D$=$\omega_B$=$\Omega$, and effective 1D coupling 
strengths, i.e. $g_{DD}$=$g_{BB}$=$g_{DB}$=$g$.  
We remark that the effective $1D$ coupling strength \cite{Olshanii} becomes  
${g_{1D}^{\sigma \sigma'}} =\frac{{2{\hbar ^2}{a_s^{\sigma \sigma'}}}}{{ma_ \bot ^2}}{\left( {1 - {\left|{\zeta (1/2)} \right|
{a_s^{\sigma\sigma'}}}/{{\sqrt 2 {a_ \bot }}}} \right)^{ -
1}}$, where $\zeta(x)$ denotes the Riemann zeta function at $x=1/2$. Here, the transversal length scale is ${a_ \bot } = \sqrt
{\hbar /{m{\omega _ \bot }}}$, with ${{\omega _ \bot }}$ the
frequency of the transversal confinement, while ${a_s^{\sigma\sigma'}}$ denotes the free space $s$-wave
scattering length within or between the two species. 
Recall at this point, that the miscibility/immiscibility condition in the absence of a trap reads $a^{12}_s \leq \sqrt{a^{11}_s 
a^{22}_s}$~\cite{Ao}.
The latter refers to the absence or presence of phase separation between the species, and our choice of equal coupling 
strengths corresponds to the miscibility/immiscibility threshold in the above expression. 
As a first approximation and motivated by the very nature of the
DB state (which is somewhat intrinsically phase separated as one
component operates as a well for the atoms of the other) and by
the insensitivity of the DB features near this threshold
in the mean-field limit~\cite{yantsitoura}, we operate at the Manakov limit
of equal values of all ${a_s^{\sigma\sigma'}}$'s. 
The interaction strength can be
tuned via ${a_s^{\sigma\sigma'}}$ with the aid of Feshbach resonances
\cite{Bloch,Chin}. Finally, note that a scaling transformation has been performed in Eq.~(\ref{Eq:9}) setting the length scale 
equal to the longitudinal characteristic oscillator length $a_{\parallel}=\sqrt{\hbar/m\omega_{\parallel}}$,  
the energy scale to $\hbar \omega_{\parallel}$ and the scaled interaction 
strength is $g=g_{1D}\sqrt{m/\hbar^3\omega_{\parallel}}$. 
The assumption of equal interactions, as also discussed in~\cite{fetter1,revip},
is tantamount to considering, e.g., the case of hyperfine
states of $^{87}$Rb; in this case also the masses are equal. 
The coefficients of interactions are not exactly equal, but are very
proximal to that limit~\cite{pe1,pe3} (see also the more recent work of Ref.~\cite{opanchuk}). 
Finally, for reasons of computational convenience, we shall set $\hbar=m=g=1$, and therefore all quantities below are given in 
dimensionless units. 

To initialize the beyond mean-field dynamics we first trace the mean-field ground state $\tilde \phi_0(x)$, 
using Newton's method. The latter is applied to the following steady-state problem:
\begin{equation}
\left[ -\frac{1}{2}\frac{d^2}{d x^2} + \frac{1}{2} \Omega^2 x^2 + | \tilde\phi_0 (x) |^2 - \mu \right] \tilde\phi_0 (x) =0.  
\end{equation}
On top of this relaxed mean-field background for fixed number of atoms $N_D$, we then embed the $N_S$ solitons 
of Eqs.~(\ref{Eq:1})-(\ref{Eq:2}) at $t=0$ with fixed amplitude as well as number of atoms $N_B$ for the 
bright component. Then the inverse widths $d_i$ are obtained using Eq.~(\ref{Eq:3}). 
We also note that the corresponding free parameters are the velocity $u_i$ 
and the position $x^0_j$ of the DB solitons. We remark that by following the above-mentioned
procedure we minimize the sound wave emission during the dynamics.  
For more details on the 
selection of the soliton and background density parameters 
we refer the reader to Appendix \ref{sec:numerics}. 

Given our interest in the qualitative characteristics of the
emerging structures,
we study their persistence with increasing 
number of allowed modes of entanglement
[specified by the number of  species functions $M$ in the Eq.~(\ref{Eq:5})] for fixed number of single particle configurations 
[specified by the number of SPFs for
species $D$, $m_D$ and $B$, $m_B$, see Eq.~(\ref{Eq:5})]. In this spirit we consider the dynamics within a ML-MCTDHB derived 
time-dependent variational approximation defined by the parameters of the ML-MCTDHB ansatz [see Eqs.~(\ref{Eq:5}), (\ref{Eq:6})] as 
$M$-$(m_D,m_B)$. Note that the $1$-$(1,1)$ approximation corresponds to the mean field case [see also Eqs. (\ref{Eq:4}), (\ref{Eq:5}), (\ref{Eq:6})].    
For further discussion on convergence issues we refer the reader to Appendix~\ref{sec:numerics1}. 

\section{Single DB soliton dynamics}

First we explore the 
oscillation of a DB solitary wave inside the parabolic trap.
We consider a binary bosonic gas consisting of $N_D=300$ atoms 
and $N_B=5$ atoms  
confined in a 1D harmonic trap with
frequency $\Omega=0.1$. The dynamics is initialized with a single DB soliton with velocity $u_1/c=0.5$ and initial position 
$x^0=-2.5$. The chemical potential of the 
background density, the amplitude and the inverse width of the DB soliton are chosen as $\mu=6.42$, $\eta=1.88$ and $d=1.42$, 
respectively (for details see Appendix A).  

\begin{figure}[t]
\includegraphics[width=0.5\textwidth]{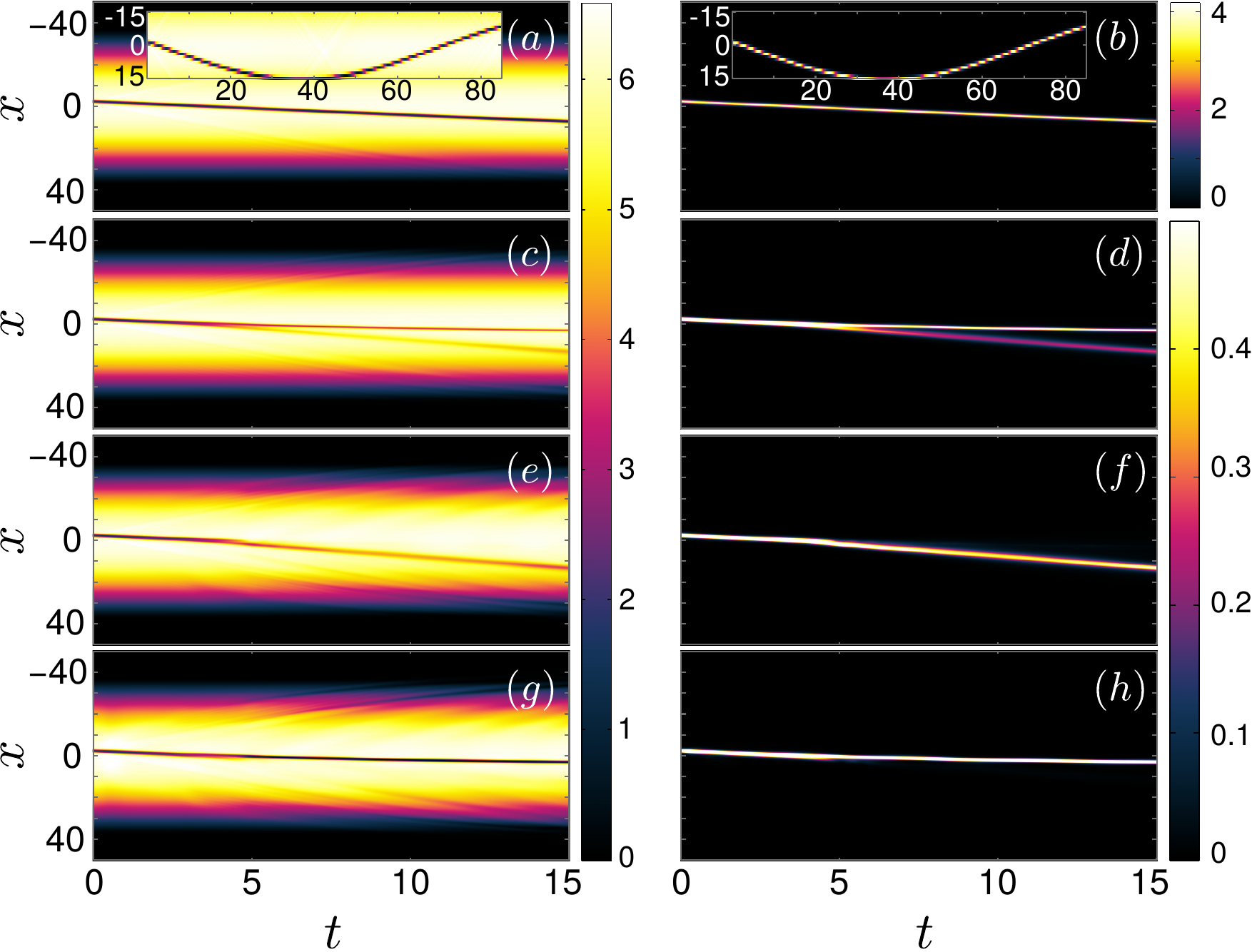}
\caption{(Color online) $\rho^{(1)}(x;t)$ of a single DB solitary wave for the 
($a$) dark and ($b$) bright component, obtained via the mean-field, i.e. 1-(1,1), approximation. The corresponding insets 
illustrate a longer time evolution to demonstrate the oscillation of the DB within the mean-field approximation.
($c$), ($d$) The same as $(a)$ and $(b)$, but 
within the correlated 15-(2,4) approximation. 
Evolution of the one-body density of the dominant species function $\Psi^{\sigma}_1$ for the ($e$) dark and ($f$) bright 
component.  
The same as before but for the next-to-dominant 
species function $\Psi^{\sigma}_2$ for the ($g$) dark and ($h$) bright component. 
The species $D$ and $B$ contain $N_D=300$ and, $N_B=5$ atoms respectively 
while the trapping frequency and the background chemical potential 
correspond to $\Omega=0.1$, and $\mu=6.42$ respectively. 
The velocity, the inverse width and the amplitude of the DB soliton 
are $u_1/c=0.5$, $d=1.42$ and $\eta=1.88$ 
respectively.} \label{Fig:1}
\end{figure}

In the mean-field case the numerically obtained (quarter-)period of oscillation for the single DB soliton is 
measured as $T_{osc}/4\approx 32.5$ (see Fig. \ref{Fig:1} ($a$), 
($b$) and the corresponding insets). This result is in good agreement with the analytical 
predictions of Ref.~\cite{buschanglin} (see also Ref.~\cite{pe3}) 
which for the parameters used herein reads $T_{osc}/4\approx 30$.
In contrast to that, in the correlated 15-(2,4) case, shown in panels ($c$) and 
($d$) of Fig.~\ref{Fig:1}, the single DB soliton decays ($t \approx 5$) into a faster,  
and a slower DB soliton (alias soliton fragments). 
\begin{figure}[ht]
\includegraphics[width=0.48\textwidth]{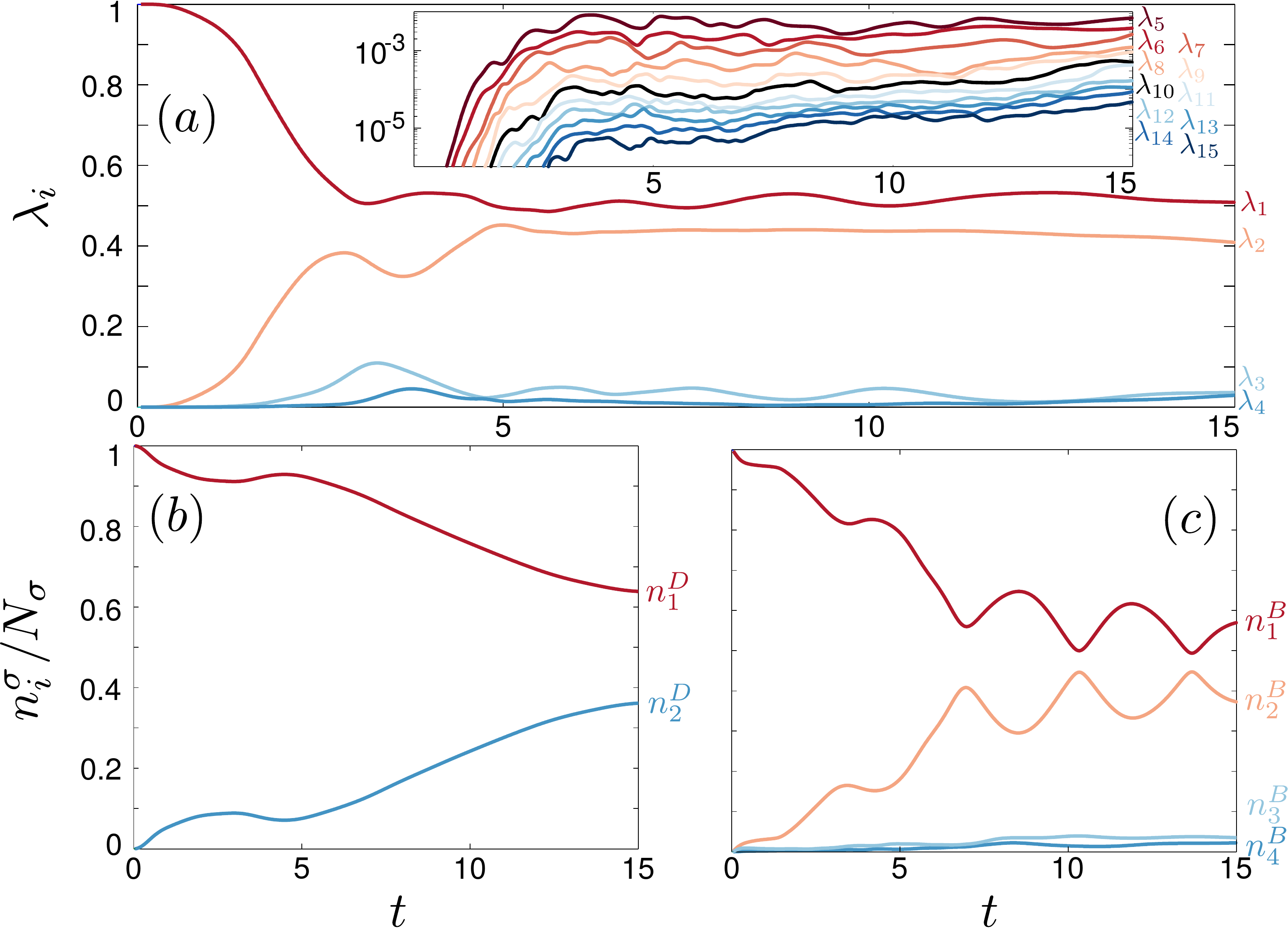}
\caption{(Color online) Evolution of ($a$) the natural occupations $\lambda_i(t)$, and the natural populations $n_i(t)$ 
for the ($b$) dark and ($c$) bright component of a DB solitary wave. The inset shows on a logarithmic scale the evolution of the 
higher-lying natural occupations 
$\lambda_5(t)$ to $\lambda_{15}(t)$. 
The parameters values are the same as in Fig. \ref{Fig:1}.}
 \label{Fig:2}
\end{figure}
%%%%%
\begin{figure*}[ht]
\includegraphics[width=1.0\textwidth]{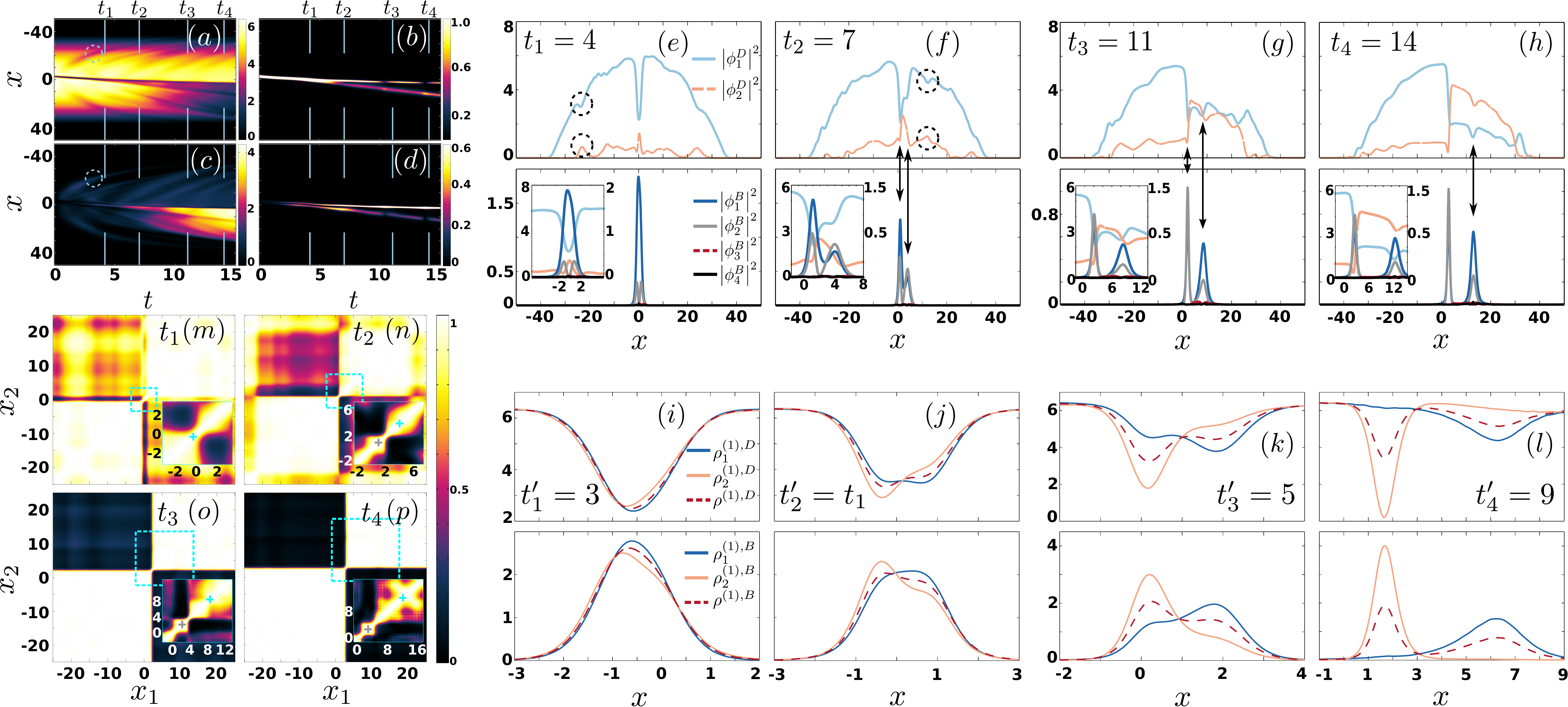}
\caption{(Color online) Density evolution of the first natural orbital for ($a$) the dark and ($b$) the bright component. 
($c$), ($d$) The same but for the second natural orbital. We remark here, that since the DB soliton is initialized at $x^0=-2.5$, the reflection symmetry is broken.  
($e$), ($f$), ($g$), ($h$) Profiles of the natural orbitals (see legend) at different time instants ($e$) $t_1=4.0$, ($f$) 
$t_2=7.0$, ($g$) $t_3=11.0$ and ($h$) $t_4=14.0$. 
For better visibility of the bright component of the DB soliton we show the insets which zoom on the position axis. 
Dashed red and solid black lines correspond to the third and fourth natural orbitals in the bright component that have nearly vanishing density population.  
The system parameters are the same as in Fig. \ref{Fig:1}. 
($i$), ($j$), ($k$), ($l$) In the upper [lower] panels
  the density profiles are shown  for
  the two most dominant dark [bright] species functions 
$\rho_1^{(1),D}(t)$, $\rho_2^{(1),D}(t)$ [$\rho_1^{(1),B}(t)$, $\rho_2^{(1),B}(t)$] and the total density 
$\rho^{(1),D}(t)$ [$\rho^{(1),B}(t)$] for different time instants during the evolution (see legend).  
($m$), ($n$), ($o$), ($p$) One-body coherence function $g^{(1),D}(x,x';t)$ for different time instants (see legend) during the evolution. 
Insets: The corresponding $g^{(1),B}(x,x';t)$ depicted within the spatial region indicated in $g^{(1),D}(x,x';t)$ by the light-blue rectangles. 
The crosses in $g^{(1),B}(x,x';t)$ refer to the position of the two solitary fragments after splitting, i.e. blue (grey) crosses indicate the 
fast (slow) DB fragment.} \label{Fig:3}
\end{figure*}
%%%%%%
By inspecting the decomposition of the total wavefunction $\Psi_{MB}$ into different modes (species functions, 
$\Psi_k^{\sigma}$), we 
observe that only two such modes are occupied significantly, during the dynamics 
[see the relevant coefficients $\lambda_1(t)$ and $\lambda_2(t)$ in Fig. \ref{Fig:2} ($a$)]. The initially 
single-mode wavefunction, quickly becomes bimodal and at the time of the decay ($t \approx 5$), both of the modes are 
characterized by almost equal occupation. 
The latter indicates the tendency of the many-body state $\Psi_{MB}(\vec x^D,\vec x^B;t)$ to approach a Schr{\"o}dinger cat state.  
After the decay, the amplitude of those modes becomes nearly constant over time. The dominant mode, $\Psi^{\sigma}_1$, 
can be identified with the fast DB soliton 
[see Fig.~\ref{Fig:1} ($e$), ($f$)] moving towards the periphery of the cloud,
while the next-to-dominant mode, $\Psi^{\sigma}_2$, is associated with the slow DB soliton 
being closer to the center of the trap [see Fig.~\ref{Fig:1}($g$),($h$)]. We remark 
here that the presented decay dynamics is in contrast to the species mean-field case, i.e. uncorrelated 
components, where the formation of the solitonic fragments is not observed [see also Fig.~\ref{Fig:11} $(e)$ in Appendix B]. 

To gain more insight into the decay dynamics of a single DB soliton we next study the population of the natural orbitals. 
We remind the reader that a state with $n_i(t)=1$ is referred to as fully condensed, while for $n_i(t)\neq 1$  
fragmentation phenomena arise~\cite{Mueller,Penrose}.
The occupations of the two natural
orbitals used for species $D$ are significant from the early times of the dynamics [see Fig. \ref{Fig:2} ($b$)].  
At times $0<t<2$, DB soliton structures appear in the first and second orbital in both species 
[see Fig.~\ref{Fig:3} ($a$)-($d$)]. 
At later times ($2<t\lesssim 4$), and as far as the dark component is concerned,
a density accumulation is observed in the second natural orbital [see Fig. \ref{Fig:3} ($c$), 
and also dashed orange line around the center of the trap in Fig.~\ref{Fig:3} ($e$)]. 
After this initial relaxation stage, the splitting
of the dark soliton leads to the emergence of two solitons manifested by
density dips appearing in the first orbital 
[see also solid ciel line at $t_2=7$ and around the center of the trap in Fig.~\ref{Fig:3} ($f$)]. 
Interestingly enough, within the cloud
and in particular around $x=-20$ for evolution times up to $t=3$,
regions with additional density depletion for
the first orbital and a corresponding density accumulation in the second orbital can also be observed,
indicated by dashed blue circles in Fig.~\ref{Fig:3}($a$) and ($c$), respectively. 
These are reminiscent of the recently experimentally observed dark-antidark states~\cite{ionut}.
For the minority species ($B$), a similar structure to the aforementioned analysis is observed. 
Since the beginning of the dynamics two different orbitals are significantly
occupied, [see $n^B_1(t)$ and $n^B_2(t)$ in Fig.~\ref{Fig:2} ($c$)]
which during evolution perform out-of-phase oscillations. 
Notice in this case the negligible population of the higher-lying
natural orbitals used for species $B$ [namely $n^B_3(t)$ and $n^B_4(t)$ in Fig.~\ref{Fig:2} ($c$)].  
The first orbital corresponds
predominantly to the fast bright soliton
[see Fig.~\ref{Fig:3} ($b$)] 
while the second orbital corresponds chiefly
to the slow bright soliton [see Fig.~\ref{Fig:3} ($d$)]. 
%%%%
\begin{figure}[ht]
\includegraphics[width=0.48\textwidth]{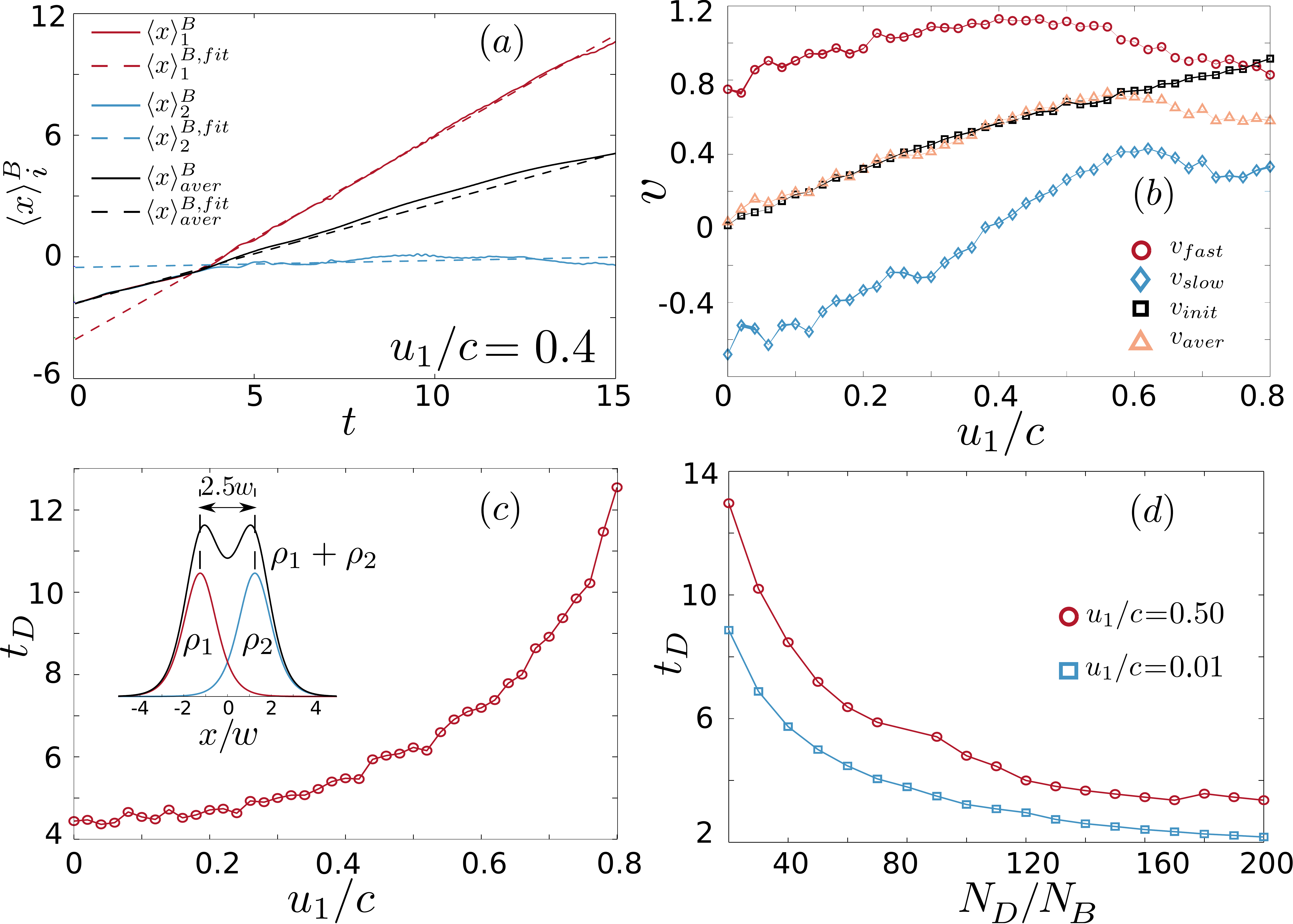}
\caption{(Color online) ($a$) Average position for each fragment that appears during the dynamics of a DB solitary wave and the 
corresponding linear fits (see legend). 
($b$) Velocity of the fast and slow fragments, the average velocity 
and the velocity after the initialization of the dynamics (see legend) for varying initial velocity. ($c$) The decay time 
$t_D$ for increasing initial velocity. 
Inset: Sketch of the calculation of the decay time via the separability of the one-body densities of the first $\rho_1$ and 
second $\rho_2$ species functions by the DB width $w=1/d$.  
($d$) The decay time $t_D$ as a function of the particle number imbalance $N_D/N_B$ with fixed particle number of the bright component ($N_B=5$) for the case of fast and slow 
velocities. These latter results have been obtained within the 15-(2,3) approximation. } \label{Fig:4}
\end{figure}
%%%%%
To further demonstrate the observed dynamics, profile snapshots of the modulus of the natural orbitals, showing some key aspects 
during the dynamics, are depicted in Fig.~\ref{Fig:3} ($e$)-$(h)$. The decay of
the DB soliton is initiated by the occupation of the second orbital of species $B$ which at initial times 
develops two density humps [see solid gray line in the bottom panel of Fig.~\ref{Fig:3} ($e$) or the relevant inset].
In species $D$, the first orbital maintains a density dip which
creates an effective potential well trapping atoms in the second
orbital of the same species [see dashed orange line in Fig.~\ref{Fig:3} ($e$) around the center of the trap]. 
As the second orbital of this species builds up, complementary patterns
reminiscent of dark-antidark solitary waves~\cite{ionut}
appear, e.g., in the same panel and around $x=-20$, or 
at later times and e.g., around $x=10$ in Fig.~\ref{Fig:3} $(f)$.
In both of the above cases the location of the formation of these matter wave 
entities is denoted by dashed black circles. 
Furthermore, in this latter panel, Fig.~\ref{Fig:3} $(f)$, and in particular 
around $x=3$ a multi-orbital DB entity is formed within the region indicated by black arrows.
The location of the formation of this structure
corresponds also to the location where the two orbitals of species $B$
both being bright solitons are aligned.  
It is also important to comment here that the two bright soliton fragments
of the first orbital (denoted by solid blue lines) are in-phase 
while the respective bright solitons formed in the second orbital of the minority 
species are out-of-phase (see nodal structures denoted by solid gray lines in the same panel).
Intriguingly, at later times and as far as the slow solitons are concerned, 
the initial dark-antidark structure evolves into 
a DW between the first and the second orbital 
of the majority species $D$ [see around the center of the trap for $t_3=11$ in Fig.~\ref{Fig:3} ($g$) also 
indicated by a black arrow].
The formation of this DW corresponds to the location
where species $B$ localizes both of its orbitals at this time.
Hence, an unusual domain-wall-bright (DWB) soliton pattern appears to arise.
Notice that at the same time, the fast DB structure is also visible around $x\approx9$ (indicated by black arrow).
For larger propagation times  
the aforementioned DWB structure remains robust 
[see top and bottom panels of Fig.~\ref{Fig:3} ($h$)],
while the multi-orbital fast DB structure is highly pronounced as indicated by a black arrow in the same figure. 

Let us now elaborate in more detail on the decay mechanism.   
As stated above, see Figs. \ref{Fig:1} ($e$)-($h$), the splitting dynamics, i.e. the decay of the initial mean-field soliton into two distinct solitary waves, 
is more transparent on the level of the species functions.  
It is already known \cite{sachadark3,sachadark2,sachadark1,sacha,sven} that a single mean-field dark soliton, when exposed to quantum fluctuations, decays due to the filling 
of the initially density depleted region by localized states.  
In our case, similar dynamics can be observed in the density evolution of the first and second natural orbital of the dark component, see Fig. \ref{Fig:3} ($e$).  
Here, the filling process progresses very rapidly as Fig. \ref{Fig:2} ($b$) suggests, in particular see the curvature of $n_1(t)$, $n_2(t)$ at the
initial time instants.   
Next, we discuss how the above-mentioned established dynamics triggers the splitting mechanism in our multispecies case by inspecting the  
species function dynamics being directly related to the natural orbitals, see also Eq. (\ref{Eq:6}).  
At times before the manifestation of the splitting the first two species functions $\Psi_i^{D}(t)$ acquire similar occupations. 
However due to their mutual orthogonality they possess different weights [i.e. $c_{1,(n_1,N-n_1)}(t)\neq c_{2,(n_1,N-n_1)}(t)$ in Eq. (\ref{Eq:6})] on the two occupied natural orbitals.   
The latter is manifested in the upper panel of Fig. \ref{Fig:3} ($i$) as a shift in the position of the corresponding density minima of $\rho_1^{(1),D}(t)$ and $\rho_2^{(1),D}(t)$, while  
the total density does not show any signatures for the subsequent splitting.  
For times close to the decay, see Figs. \ref{Fig:3} ($j$), ($k$), the shifting between $\rho_1^{(1),D}(t)$ and $\rho_2^{(1),D}(t)$ becomes more prominent 
and each $\rho_i^{(1),D}(t)$ exhibits two density minima due to the mutual coupling of $\Psi_1^{D}(t)$ and $\Psi_2^{D}(t)$.  
Furthermore, the two aforemetioned density minima are also imprinted in the total density. 
Finally, for later evolution times after the splitting [see Fig. \ref{Fig:3} ($l$)] the $\rho_i^{(1),D}$'s become progressively well separated and each one exhibits only one density minimum which 
corresponds to the two solitonic structures.
We remark here that according to the Schmidt decomposition [see Eq. \ref{Eq:5}] $\Psi_i^{B}(t)$    
is directly related to $\Psi_i^{D}(t)$. 
Indeed, as shown in the lower pannels of Figs. \ref{Fig:3} ($i$)-($l$) the density maxima of $\rho_i^{(1),B}(t)$ follow the density minima of $\rho_i^{(1),D}(t)$. 
Thus, the decay mechanism is also manifested in the bright component. 
Summarizing, the splitting mechanism is a manifestation of the dynamical build
up of the first two species functions due to the 
interspecies coupling and it is triggered by the filling of the first orbital by the second one in the dark component. 

To infer about the degree of first order coherence during the DB soliton dynamics, we employ the 
normalized spatial first order correlation function \cite{Naraschewski}   
$g^{(1),\sigma}(x,x';t)=\rho^{(1),\sigma}(x,x';t)/\sqrt{\rho^{(1),\sigma}(x;t)\rho^{(1),\sigma}(x';t)}$. 
Here, $\rho^{(1),\sigma}(x;t)$ refers to the one-body density matrix of $\sigma$ species at position $x$. 
We also note that $|g^{(1),\sigma}(x,x';t)|^2$ is bounded taking values within the range $[0,1]$. 
Indeed, a spatial region with $|g^{(1),\sigma}(x,x';t)|^2=0$ is referred to as perfectly incoherent, while if 
$|g^{(1),\sigma}(x,x';t)|^2=1$ is said to be fully coherent.  
Especially, in our case we are interested for the appearance of Mott-like correlations.
%being identified by
By this we mean the localization 
of the one-body correlations within a certain spatial region $R$ (i.e. $|g^{(1),\sigma}(x,x';t)|^2 \approx 1$ $x,x'\in R$) and perfect incoherence between different 
spatial regions $R$, $R'$ (i.e. $|g^{(1),\sigma}(x,x';t)|^2\approx 0$, $x\in R$, $x'\in R'$ with $R \cap R' = \varnothing$). 
Figs. \ref{Fig:3} ($m$)-($p$) present $|g^{(1),D}(x,x';t)|^2$ for different time instants during the dynamics, namely before and after the splitting 
of the initial mean-field DB soliton. 
As it can be seen, there is a gradual formation of Mott-like correlations, in particular after the decay, being separated by the location where the DWB develops. 
Similar type of correlations is also observed for the bright component, see the insets in Figs. \ref{Fig:3} ($m$)-($p$). 
Indeed, a localization of the $|g^{(1),B}(x,x';t)|^2$ is manifested within the spatial region in which the density of the DWB is dominant. 

In order to examine the validity of a form of the particle picture (which is known to hold in the mean-field limit) 
in terms of the above-mentioned solitary 
fragments of the initially imprinted DB soliton, we employ the position expectation value of the first 
(corresponding to the fast DB structure) and the second (attributed to the slow DB structure respectively)
species function that reads
\begin{equation}
\langle x \rangle_i^{B}(t)=\frac{1}{N_B} \int^{\infty}_{-\infty} dx~ x \rho_i^{(1),B}(x;t).
\end{equation}
Here, $N_B=\int^{\infty}_{-\infty} dx~ \rho_i^{(1),B}(x;t)$ refers to the particle number  
and $\rho_i^{(1),B}(x;t)$ denotes 
the one-body density of the $i$-th $B$ species function. 
We shall show that the aforementioned decay process results into two DB solitons possessing the same mass,
while the momentum of the initial DB ``particle'' is transferred to the resulting DB fragments
so as for the total momentum to be conserved. To this end, we also calculate the average mean position 
i.e. $\langle x \rangle_{aver}^B=\frac{1}{2}(\langle x \rangle_{1}^B+\langle x \rangle_{2}^B$).  
Fig.~\ref{Fig:4}($a$) illustrates
the position expectation value of the first or the 
second species function and their corresponding fits, as well
as the average position obtained from these two. The  
linear fits to $\langle x \rangle_{1}^B$, $\langle x \rangle_{2}^B$ and $\langle x \rangle_{aver}^B$ are shown by dashed lines. 
For  $t<t_0$ (no splitting of the DB soliton)
the linear extrapolation of  $\langle x \rangle_{1}^B$, $\langle x \rangle_{2}^B$
is also shown, where the time instant 
$t_0$ is identified by the condition $\langle x \rangle_{1}^B \backsimeq \langle x \rangle_{2}^B \backsimeq \langle x \rangle_{aver}^B$.   
The fitting is chosen to reflect
the time for which different entities exist.
The observed deviations $\langle x \rangle _i^B-\langle x \rangle _i^{B,fit}$ 
are small, while the deviation $\langle x \rangle _{aver}^B-\langle x \rangle _{aver}^{B,fit}$
is larger. The latter can be attributed to the coupling of the first two species functions with the remaining of the species functions. 
Among the two, the somewhat more significant contribution to $\langle x \rangle _{aver}^B-\langle x \rangle _{aver}^{B,fit}$ stems from 
$\langle x \rangle _{2}^B-\langle x \rangle _{2}^{B,fit}$ 
suggesting that the second species functions possesses a stronger coupling with the remaining species functions. 

\begin{figure}[ht]
\includegraphics[width=0.48\textwidth]{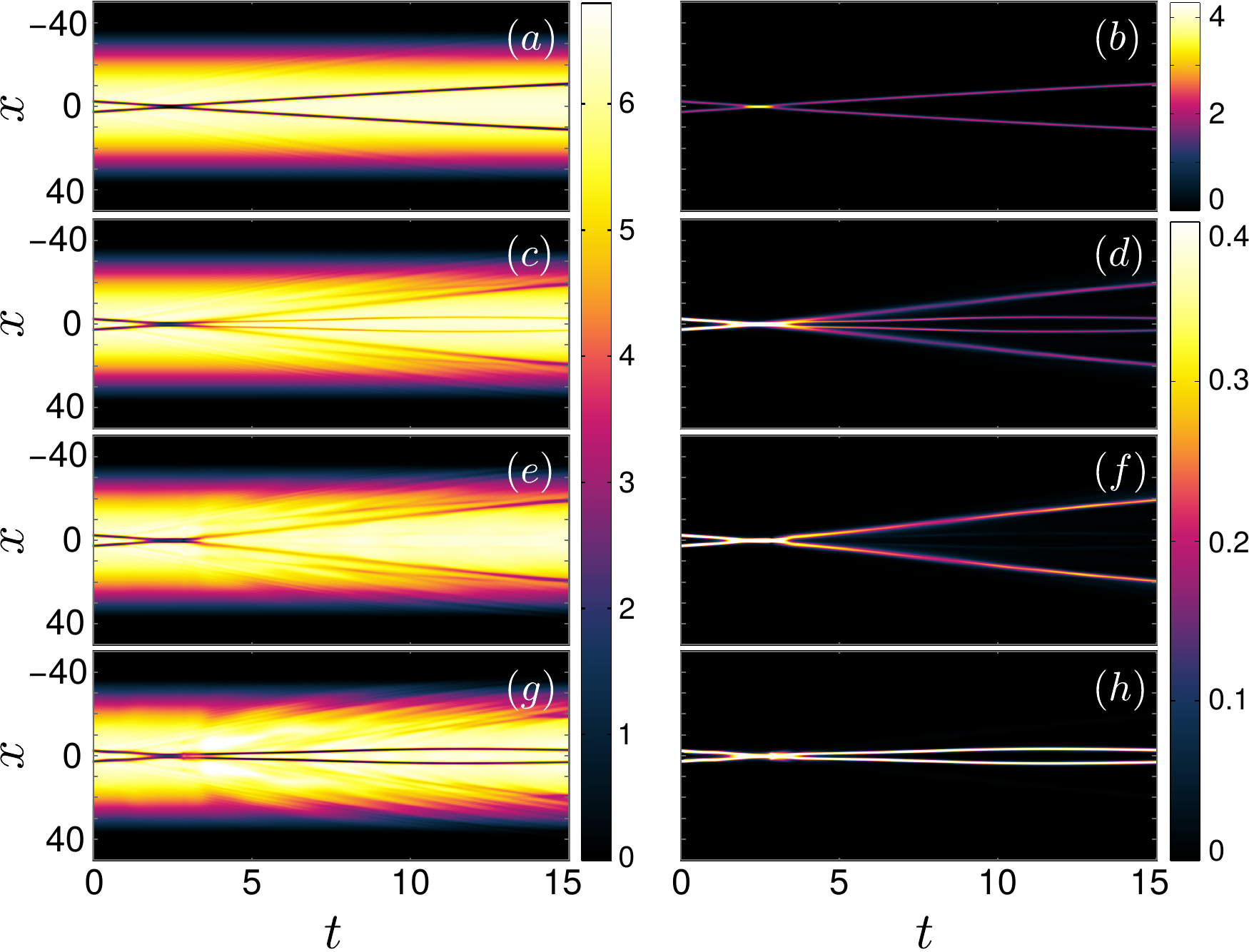}
\caption{(Color online) One-body density evolution $\rho^{(1)}(x;t)$ for the ($a$) dark and ($b$) the bright component 
obtained via the mean-field, i.e. 1-(1,1), approximation. 
($c$), ($d$) The same as above but within the correlated 15-(2,4) approximation.
Density evolution of the dominant species function $\Psi^{\sigma}_1$ for the ($e$) dark and ($f$) bright component.  
The same as before but for the next-to-dominant species function $\Psi^{\sigma}_2$ for the ($g$) dark and ($h$) bright 
component. 
%Each 
The species contains $N_D=300$, $N_B=5$ atoms while the trapping frequency and the background chemical potential 
correspond 
to $\Omega=0.1$, and $\mu=6.47$ respectively. 
The velocity, the inverse widths, the amplitudes and the initial positions of the DB solitons correspond to $u_1/c=u_2/c=0.5$, 
$d_1=d_2=1.82$, $\eta_1=\eta_2=1.51$ and $x^0_1=-2.5=-x^0_2$ respectively.} \label{Fig:5}
\end{figure}

\begin{figure}[ht]
\includegraphics[width=0.45\textwidth]{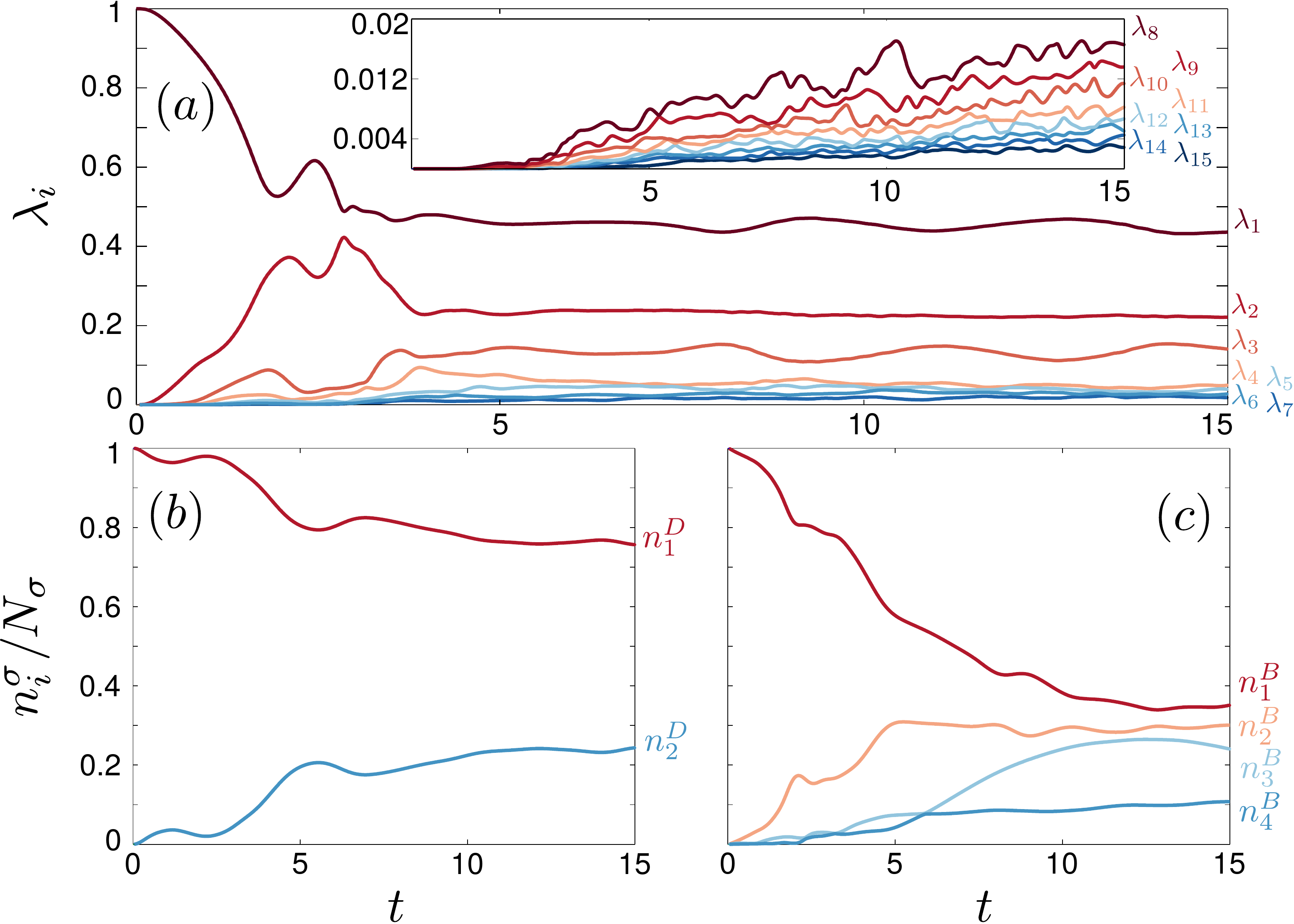}
\caption{(Color online) Evolution of ($a$) the natural occupations $\lambda_i(t)$, and the natural populations $n_i(t)$ 
for the ($b$) dark and ($c$) bright component of two colliding DB solitary waves. 
The inset shows in a magnified scale the evolution of the higher-lying natural 
occupations $\lambda_8(t)$ to $\lambda_{15}(t)$. 
The %system 
parameter values are the same as in Fig. \ref{Fig:5}.} \label{Fig:6}
\end{figure}

\begin{figure}[ht]
\includegraphics[width=0.42\textwidth]{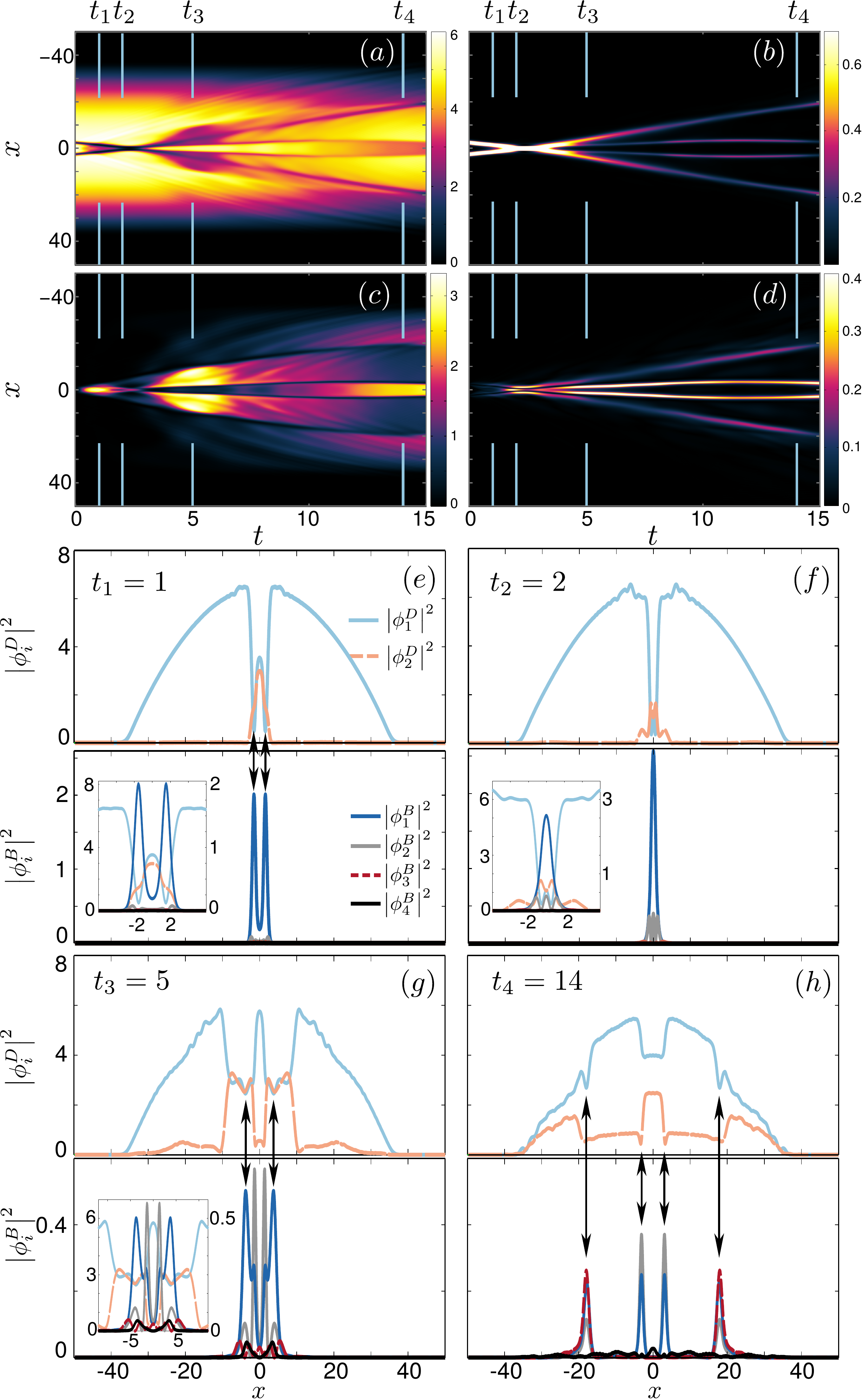}
\caption{(Color online) Density evolution of the first natural orbital for the ($a$) dark and ($b$) the bright component. 
($c$), ($d$) The same as before but for the second natural orbital. 
($e$), ($f$), ($g$), ($h$) Profiles of the natural orbitals (see legend) at different time instants ($e$) $t_1=1.0$, ($f$) 
$t_2=2.0$, ($g$) $t_3=5.0$ and ($h$) $t_4=14.0$. 
For better visibility of the bright component of the two DB solitons we show the insets which 
zoom into the center of the trap. 
The %system 
parameter values are the same as in Fig. \ref{Fig:5}.} \label{Fig:7}
\end{figure}

%%%%
Following the above-described procedure for the range of initial velocities $u_1/c\in \{0, 0.8\}$ we show in Fig. \ref{Fig:4}($b$) 
the resulting velocities for the 
fast $v_{fast}$ (being the slope of $\langle x \rangle_{1}^{B,fit}$) 
and slow $v_{slow}$ (being the slope of $\langle x \rangle_{2}^{B,fit}$)
fragments of the bright component (see also Fig. \ref{Fig:1} ($f$), ($h$) 
respectively). To discuss whether particle picture considerations
are helpful in this alternative 
framework we also present the imprinted velocities after the initialization of the 
dynamics, denoted by $v_{init}$ \cite{note_vel} (i.e. the slope of $\langle x \rangle_{aver}^{B,fit}$) 
and the corresponding average velocity $v_{aver}=\frac{v_{fast}+v_{slow}}{2}$. It is observed 
that for $u_1/c<0.6$ the particle picture 
makes sense, as $v_{aver}=\frac{v_{fast}+v_{slow}}{2}=v_{init}$, indicating the conservation of momentum. In this case, 
each component of the DBs splits into two fragments 
possessing the same mass [see also the relevant occupations in Fig.~\ref{Fig:2} $(a)$]. However, for $u_1/c>0.6$  
larger deviations from the particle picture are evident. This can be attributed to the fact that one of the emergent 
fragments within the DB entity is ``heavier" from the other one due to a rapid mass redistribution caused by the large 
velocities. 

Next, let us examine how the decay time $t_D$ observed during the dynamics depends on both the initial velocity of the DB structure  
and the particle number of the dark 
component ($N_D$). To quantitatively consider the decay time we impose the criterion $\langle x \rangle_1^B-\langle x \rangle_2^B>2.5w$, where $w=1/d$ refers to the  
width of the DBs [see also the inset of Fig.~\ref{Fig:4} ($c$)]. As it can be seen, in Fig.~\ref{Fig:4}($c$), 
the decay time shows a gradual growth 
for increasing initial velocity, indicating that in the limit of large velocities we tend to the mean-field picture.  
On the other hand, Fig.~\ref{Fig:4}($d$) presents the 
corresponding decay time $t_D$ for different particle number imbalances $N_D/N_B$ (here defined by varying the $N_D$ and keeping $N_B=5$ fixed) for the case of fast ($u_1/c=0.5$) and 
slow ($u_1/c=0.01$) velocities. 
An overall decrease of the decay time for increasing particle number $N_D$ is observed both for slow and fast solitons. 
This can be intuitively understood from the fact that a larger imbalance ($N_D/N_B$)  ``effectively'' corresponds to a smaller bright filling.   
The latter implies that the DBs are more stable than dark solitons (limit $N_D/N_B \to \infty$) under beyond mean-field correlations.  
Finally, as expected [see also Fig.~\ref{Fig:4}($c$)] 
the corresponding decay times are larger for fast solitons than for slow ones. 

Let us also remark here that our results of Figs. \ref{Fig:4} ($b$), ($c$) in conjuction with the decay mechanism presented before 
[see also Figs. \ref{Fig:3} ($i$)-($l$)] suggest that the resulting solitonic structures do not undergo  
further splittings at least within the presented evolution time scales.
On the one hand, the slow velocity fragment is not a mean-field DB and therefore the observed 
splitting process of a purely mean-field DB is not expected to occur for such a structure. 
The slow velocity fragment evolves from dark-antidark and bright states to a DWB. 
The latter possesses no mean-field analogue for a binary system and its stability is presently not known; this constitutes a particularly interesting
topic for future study. 
However, it is expected to be robust as it is a correlated structure produced by quantum fluctuations, and therefore it is not likely 
that the same mechanism can lead to its decay.  
In that light, we interpret our results as an indication of stability of this beyond mean-field structure.  
On the other hand, the fast soliton fragment is more proximal to a mean-field solitary wave and as such it is expected to be 
more prone to decay.  
Fig. \ref{Fig:4} ($b$) suggests a velocity $u_1/c>0.7$ which in combination to 
Fig. \ref{Fig:4} ($c$) shows a decay time $t_D>8$, being     
much larger than the decay time of the parent mean-field soliton ($t_D\approx 5$). 
Therefore even if a splitting process is permitted it can not be observed 
within the presented evolution times. 
Note here that the fast solitary fragment is fully formed only at times $t>11$, see Figs. \ref{Fig:3} ($g$), ($h$). 
However, we have to be cautious since this fragment is also not of a pure mean-field nature, and therefore its stability is 
highly non-trivial.

\section{Two DB soliton dynamics}

We now turn to the examination of two-soliton dynamics involving
collisions. In particular, both fast collisions where the solitons
rapidly go through each other (dominated by kinetic energy), 
as well as slow collisions where the inter-particle interaction
(and the potential
energy landscape) dictate the dynamics are investigated. 
Furthermore, in all cases to be presented below, the two DB solitons have common inverse widths, and are initialized at
$x^0_1=-x^0_2=-2.5$ travelling towards one another with velocity, $u_1/c=-u_2/c$. 

\subsection{Fast Soliton Collisions}

The results for the case of the fast solitons are summarized in Figs.~\ref{Fig:5}-\ref{Fig:7}. It can be seen that
the mean-field picture differs substantially from the beyond-mean-field one. 
The reason for this is that while the mean-field scenario displays simply the collision of the two 
DBs [see Fig.~\ref{Fig:5} $(a)$ and $(b)$], the
beyond-mean-field case features not two, but demonstrably (at least) four DBs, two of which can be thought of as outer ones and two 
of which are inner ones [see Fig.~\ref{Fig:5} $(c)$ and $(d)$]. 
However, and upon examining the dominant $\Psi^{\sigma}_1$ [see Fig.~\ref{Fig:5} $(e)$ and $(f)$]
and the next-to-dominant order $\Psi^{\sigma}_2$ species functions [see Fig.~\ref{Fig:5} $(g)$ and $(h)$]
for both the dark and the bright solitons respectively,
a clearer picture of the resulting dynamics can be drawn.
In particular, during the first stage and the soliton collision ($t\approx 2$)
the second mode (i.e. $k=2$ in Eq. (\ref{Eq:5})) becomes highly populated, and at later times, predominantly features the inner DBs 
[see Fig.~\ref{Fig:5} $(g)$ and $(h)$].
At the same time, the first mode of entanglement chiefly consists of the outer DBs [see Fig.~\ref{Fig:5} $(e)$ and $(f)$], 
although it also supports a weaker DB-like feature at the location of the inner DB structure. 
The latter trait indicates that perhaps the inner DBs can be thought 
of as featuring multiple modes of entanglement. The aforementioned dynamics is 
further captured by a more careful inspection of the 
natural occupations of the species 
functions [see Fig.~\ref{Fig:6} $(a)$].
It becomes evident that the collision occurring at earlier times 
entails a rapid population redistribution stage. One can clearly infer that the different modes of entanglement 
acquire comparable occupations which subsequently become roughly constant throughout the propagation. 
Notice also the negligible population of the species functions $\lambda_8(t)$ and higher, 
shown in the inset of Fig.~\ref{Fig:6}$(a)$. 
To further demonstrate the deviations from the mean-field approximation,
the evolution of the natural populations is employed [see Fig.~\ref{Fig:6}$(b)$ and $(c)$]. 
In particular a rapid depletion is observed 
up to the collision event for both the dark and the bright soliton components.
The fragmentation process is indicated by the decrease of population in the first orbital and the simultaneous 
emergence of population in the second natural
orbital (and in all four orbitals) for the majority species $D$ (minority species B). 
Closely following the space-time evolution of each of the natural orbitals 
of the majority species $D$
[see Fig.~\ref{Fig:7} $(a)$ and $(c)$]
the existence of a density depletion in the first orbital, with an associated  
density accumulation in the second orbital is observed. 
Similarly the first two orbitals associated with the
minority species $B$ feature bright solitons in both locations, although arguably the first one is (in comparison with
the second) slightly more populated at the outer bright, while the second one is definitely more
populated (in comparison with the first) at the inner one 
[see Fig.~\ref{Fig:7} $(b)$ and $(d)$]. 
Moreover, since the picture in the context of the orbitals is somewhat less transparent,
in panels $(e)$-$(h)$ of Fig.~\ref{Fig:7}, profile snapshots of the density evolution of each of the aforementioned natural 
orbitals is depicted.  
At initial times [see $t_1=1$ in Fig.~\ref{Fig:7} $(e)$], and before the collision takes place, the second orbital of the 
majority species develops a bright structure [see dashed orange line in panel $(e)$] within the region at the center [see solid ciel line in panel $(e)$]. 
The latter, seemingly, develops into two anti-dark solitons [see $t_2=2$ in 
Fig.~\ref{Fig:7} $(f)$ top] with the respective bright counterpart [see $t_2=2$ in Fig.~\ref{Fig:7} $(f)$ bottom] 
featuring a particle concentration aligned with the merger of the two initial bright solitons, at the collision 
point. Just after the collision multiple DWB structures are realized for the inner DB solitary wave 
[see orange and gray dashed and solid lines respectively $t_3=5$ in Fig.~\ref{Fig:7} 
$(g)$]. These persist for larger propagation times [see inner solitonic structure denoted by black arrows 
at $t_4=14$ in Fig.~\ref{Fig:7} 
$(h)$]. Furthermore, the previously suppressed contribution of the third natural orbital of the minority species $B$
becomes dominant at this larger propagation times, featuring the outer DB soliton pattern [see dashed red lines in the bottom 
panel of Fig.~\ref{Fig:7} $(h)$ also indicated by black arrows]. This latter observation leads to further support of the 
multi-orbital nature of the solitonic structures.   
We also remark that the same qualitative results remain valid for larger number of atoms $N_D$ (results not shown here for brevity). 

\begin{figure}[ht]
\includegraphics[width=0.5\textwidth]{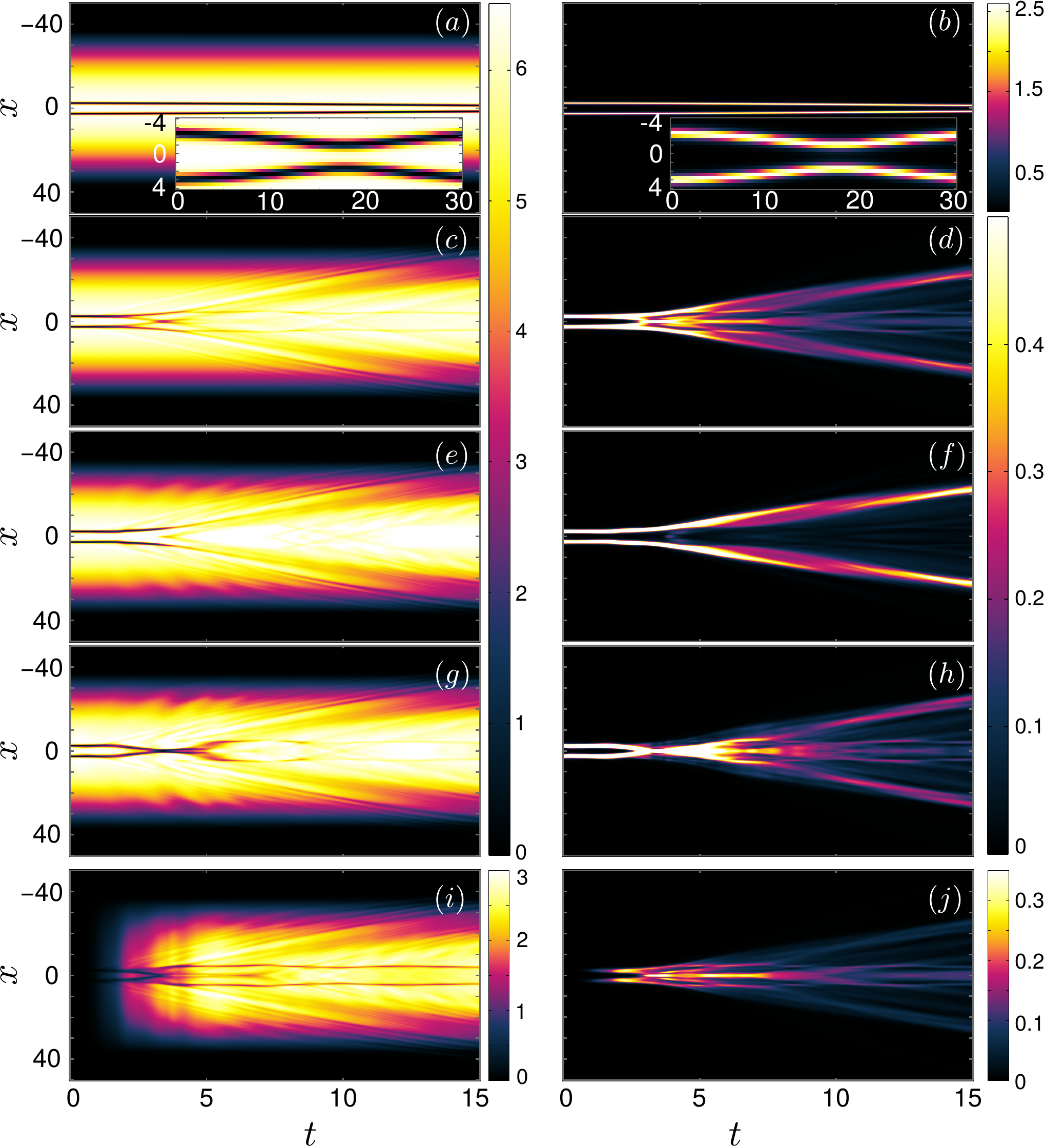}
\caption{(Color online) One-body density evolution $\rho^{(1)}(x;t)$ for the ($a$) dark and ($b$) the bright component 
obtained via the mean-field, i.e. 1-(1,1), approximation. Insets in both panels illustrate the long-time evolution of the 
respective densities, to demonstrate the repulsion between the two DB solitons. 
($c$), ($d$) The same as above but within the correlated 15-(2,4) approximation.
Density evolution of the dominant species function $\Psi^{\sigma}_1$ for the ($e$) dark and ($f$) bright component.  
The same as before but for the next-to-dominant $\Psi^{\sigma}_2$ 
[the re-summation of higher order contributions $\sum\limits_{i>2} \lambda_i 
\rho_i^{(1),S}(x)$] species function for the ($g$) [($i$)] dark 
and ($h$) [($j$)] bright component.  
%Each 
The species contain $N_D=300$, $N_B=5$ atoms while the trapping frequency and the background chemical potential correspond 
to $\Omega=0.1$, 
and $\mu=6.5$, respectively. 
The velocity, the inverse widths, the amplitudes and the initial positions of the DB solitons correspond to $u_1/c=u_2/c=0.01$, 
$d_1=d_2=2.0$,
$\eta_1=\eta_2=1.58$, and $x^0_1=-2.5=-x^0_2$ respectively.} \label{Fig:8}
\end{figure}

\begin{figure}[ht]
\includegraphics[width=0.45\textwidth]{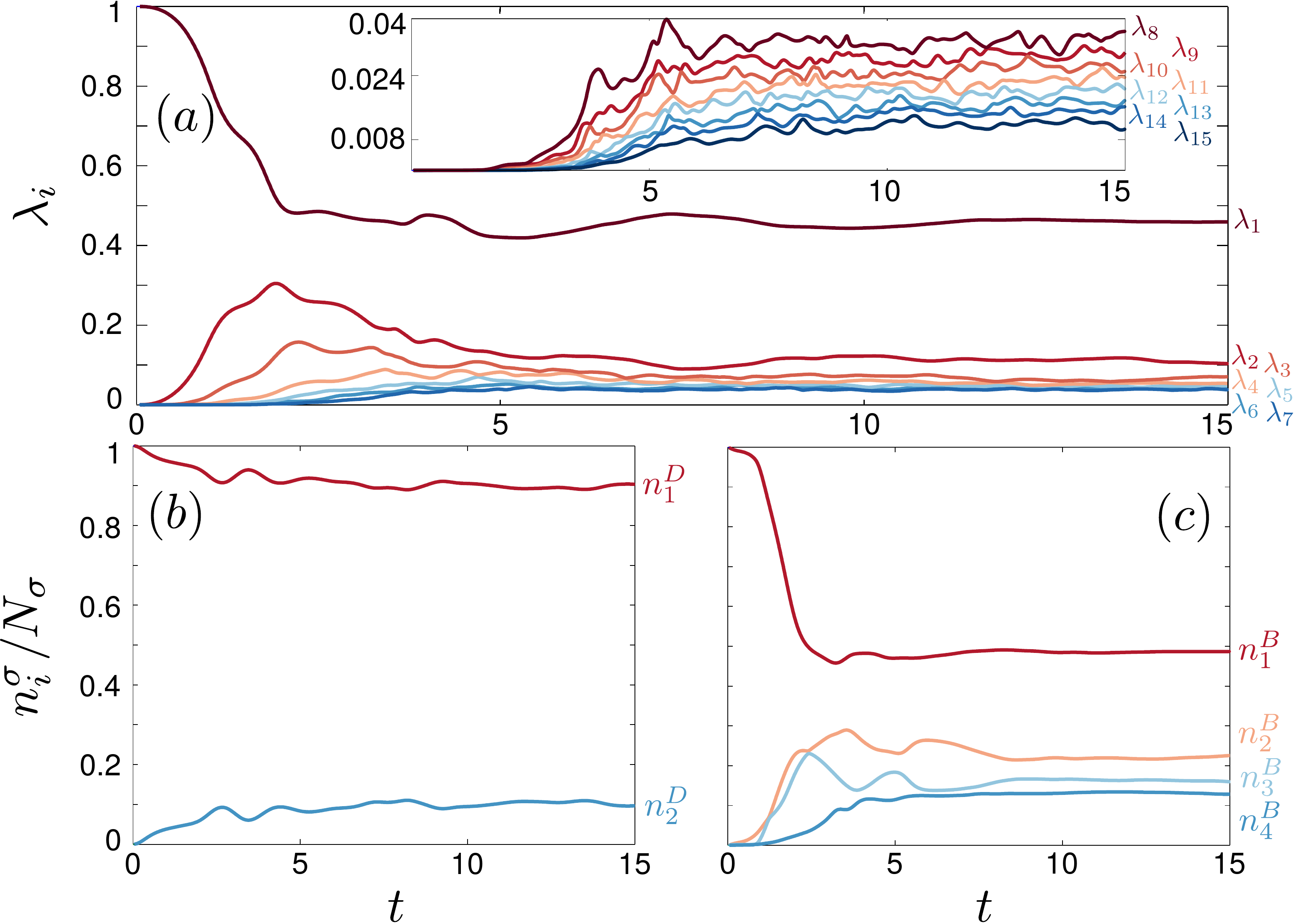}
\caption{(Color online) Evolution of ($a$) the natural occupations $\lambda_i(t)$, and the natural populations $n_i(t)$ 
for the ($b$) dark and ($c$) bright component of two colliding DBs. The inset shows on a magnified scale the 
evolution of the higher-lying natural 
occupations $\lambda_8(t)$ to $\lambda_{15}(t)$. 
%The system 
Other parameters used are the same as in Fig. \ref{Fig:8}.} \label{Fig:9}
\end{figure}

\begin{figure}[ht]
\includegraphics[width=0.5\textwidth]{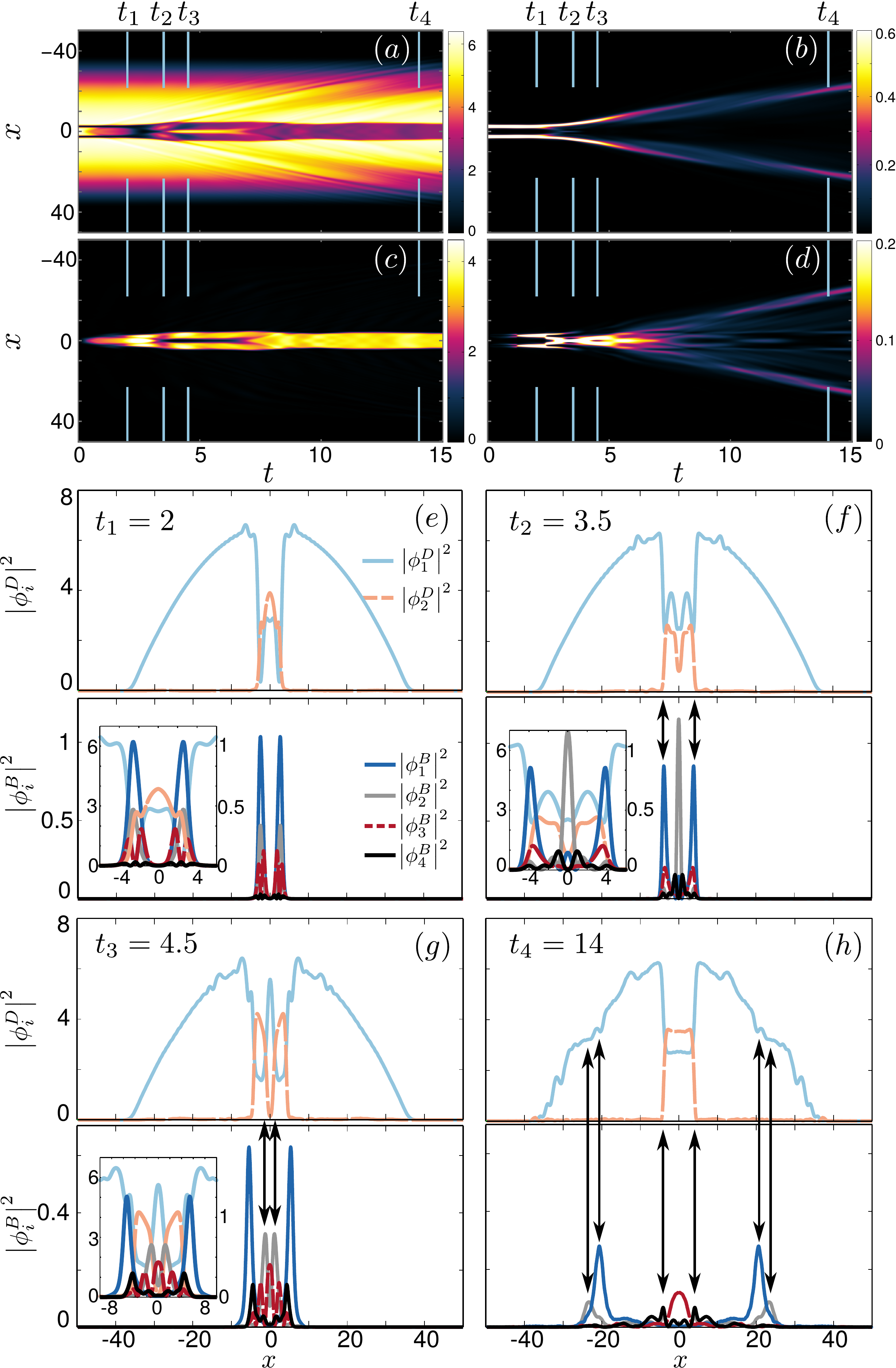}
\caption{(Color online) Density evolution of the first natural orbital for the ($a$) dark and ($b$) the bright component. 
($c$), ($d$) The same as above but for the second natural orbital. 
($e$), ($f$), ($g$), ($h$) Profiles of the natural orbitals (see legend) at different time instants ($e$) $t_1=2.0$, ($f$) 
$t_2=3.5$, ($g$) $t_3=4.5$ and ($h$) $t_4=14.0$.  
For better visibility of the bright component of the DB solitary waves, we show the insets which zoom into the center of the 
trap. Parameter values used are the same as in Fig.~\ref{Fig:8}.} \label{Fig:10}
\end{figure}

\subsection{Slow Soliton Collisions} \label{sec:convergence} 

We now turn to an example of a slow collision, summarized in Figs.~\ref{Fig:8}-\ref{Fig:10}, following a similar format 
as before. Here, at first glance and for times up to $t\approx3$ 
the dynamics of the one-body densities appears to be similar between the 
mean-field (shown in panels $(a)$ and $(b)$ of Fig.~\ref{Fig:8}) and the beyond-mean-field scenario 
(shown in panels $(c)$ and $(d)$ of Fig.~\ref{Fig:8}).
However, at later times 
and as is clearly shown in panels $(e)$-$(j)$ of Fig.~\ref{Fig:8}, 
the mean-field approximation is far from an adequate description of the dynamics. 
At the mean-field level, the in-phase dynamics of the bright components leads to an effective net repulsion~\cite{seg2,pe3} 
between them that occurs at large propagation times ($t\gtrapprox 20$)
as indicated  by the insets in 
panels $(a)$ and $(b)$ of Fig.~\ref{Fig:8}.
This repulsion along with the restoring force of the trap is essentially responsible for the observed 
phenomenology of the dynamics. On the other hand (and for $t>3$), within the beyond mean-field approximation,  
repulsion is evident at the one-body density level depicted in panels $(c)$ and $(d)$.
In particular, it can be clearly discerned that the first mode ($k=1$ in Eq.
(\ref{Eq:5})) features a DB solitonic structure which repels and moves outward [see panels $(e)$ and $(f)$], 
while the second mode [see panels $(g)$ and $(h)$] acquires a DB with demonstrable attraction.
This inner DB solitonic structure after a first collision at about $t\approx 3.5$
seems to settle at a nearly constant distance from each other in an effective ``bound state''. 
It is relevant to mention that in the single-component mean-field realm,
such a bound state
is only relevant in the case of out-of-phase bright structures in the minority component~\cite{seg2,pe3}.
To further illustrate this effective bound state, but also to reveal its multi-orbital nature, 
in panels $(i)$ and $(j)$ of Fig.~\ref{Fig:8} a summation 
over the higher species function contributions, i.e. $\sum\limits_{i>2} \lambda_i \rho_i^{(1),S}(x)$, is depicted. 
As in the case of fast soliton collisions,
a redistribution of mass is observed between 
the entangled modes illustrated in Fig.~\ref{Fig:9} $(a)$. 
Two modes are significantly contributing (possessing an occupation higher than $0.1\%$)   
during the evolution, see the corresponding coefficients $\lambda_1(t)$, and $\lambda_2(t)$.
This is in contrast to the fast collision scenario presented above,
where three modes were significantly occupied. The higher-lying species 
functions possess negligible occupations [see the inset in Fig.~\ref{Fig:9}$(a)$]. 
Furthermore, the fragmentation in this velocity regime is less pronounced for
the majority species $D$ but of about $50\%$ 
for the minority species $B$, shown in Fig.~\ref{Fig:9} $(b)$, $(c)$ respectively. 
The respective evolution of the two natural orbitals for both components is shown  
in Fig.~\ref{Fig:10} $(a)$-$(d)$. 
Notice, that once more the picture in terms of orbitals is less transparent.
However, the first orbital 
can be associated with the outer DB solitary wave. This observation is clearly evident in the bright counterpart [see Fig.~\ref{Fig:10} $(b)$] and less 
pronounced in the corresponding dark component [see Fig.~\ref{Fig:10} $(a)$].   
The excitation of the 
second orbital predominantly leads to the formation of the inner soliton. 
In the latter, traces of attractivity can be identified over time both in the gradually approaching 
dark component, as well as in the periodically merging (and separating again) bright component. 
To further elaborate on the resulting dynamics, different instants during propagation are shown in  
Fig.~\ref{Fig:10} $(e)$-$(h)$.
It is observed that initially the second natural orbital of the majority species $D$ features a density hump
located at the center of the trap. The respective four orbitals of the minority species $B$ are all aligned with 
the two dark solitons of the first natural orbital of the majority species $D$ [see $t_1=2$ top and bottom panels of 
Fig.~\ref{Fig:10} $(e)$ and the relevant inset].
Notice also that the second orbital of the minority species develops two 
bright solitons.
In particular, since the first orbital consists of in-phase bright counterparts 
the second orbital (due to orthogonality) generates out-of-phase brights. 
It is this latter effect that naturally leads to the formation of the effective ``bound state'' 
discussed above [see also Fig.~\ref{Fig:8}$(i)$-$(j)$]. We note here, that an analogous result was also reported in 
Ref.~\cite{cederbaum}.
The respective third natural orbital develops at the same time density dips.
Thus a multi-orbital DB soliton emerges.
At the  collision point, shown in Fig.~\ref{Fig:10} $(f)$, the outer DB soliton is clearly visible 
(indicated by black arrows), and is accompanied by the inner DB structure formed by the first and second natural orbitals of 
the majority species $D$ and predominantly by the second orbital of the respective bright soliton component. 
After the collision event [see $t_3=4.5$ Fig.~\ref{Fig:10} $(g)$], the outer DBs can be seen to move outwards. 
However, a more complicated picture is drawn by the higher natural orbitals.
Namely, a dark soliton formed in the second orbital of species $D$ creates an effective potential 
that traps atoms of the third orbital of species $B$ around the center of the trap,  
as shown in the inset of Fig.~\ref{Fig:10} $(g)$. 
The inner DB structure, indicated by black arrows in the same panel,
is aligned with the two dark solitons formed in the third orbital of species $B$, verifying its multi-orbital 
nature.
Finally, at later times, depicted in Fig.~\ref{Fig:10} $(h)$, both the outer and inner DBs are moving towards the 
boundaries (all four are indicated by black arrows, although
the dark solitons of the first orbital in the majority component
are less clearly
discernible).  
Furthermore, a DW structure developed by the two natural orbitals of the 
majority species traps the bright solitons formed in the fourth orbital of species $B$ around the center of the trap.
Thus, even in this case a DWB structure arises.

\section{Conclusions and Future Challenges}

In the present work, we examined the dynamics of one, as well
as of two DB solitons comparing the single-orbital
mean-field approximation with the beyond-mean-field approximation 
featuring multiple inter-species modes of entanglement (and multiple orbitals).
A predominant conclusion is that both the oscillation
and the interaction between solitons appear to lead to
their fragmentation and the population of additional modes
(and orbitals). In the typical case scenario,
the fragmentation leads to the decay of the initial DB solitons and the simultaneous 
emergence of an outer DB
pertaining to the dominant mode, and an inner
one associated chiefly with the first subdominant mode.
In the process, we could also identify more complex
patterns including, e.g., DWs even within the
orbitals of the same (majority $D$) species, as well as
patterns reminiscent of dark-antidark structures.

Furthermore, for the single DB soliton
dynamics, it has been shown that the  
considerations involving momentum redistribution at the
particle level
between the modes of entanglement capture quite accurately
the beyond mean-field
decay. Additionally the mean-field DBs were found to be more robust against decaying upon increasing  
the initial velocity. It was also found that the lifetime decreases with the particle number 
imbalance between the dark and the bright soliton components.  

In the case of collisions, as an effect of the interaction between the constituent DBs,
a more involved excitation dynamics when compared to the single DB case is observed. 
In particular it is found that both fast
and slow soliton collisions appear rather disparate from their mean-field
analogues as far as the one-body density is concerned.
In the former case, this is due to the inner DBs featuring
a multi-orbital state more pronounced at 
the one-body density level. Yet, arguably, slow collisions
are fairly dramatic too in repartitioning the density to
inner, realizing nearly a bound state, and outer solitary
waves. Particularly this bound state is created
due to the spontaneous emergence of out-of-phase bright solitons 
building up in the next-to-dominant order species function for the 
bright component. 

Let us also comment on the corresponding experimental realization of our 
setup. DB soliton structures have already been realized in \cite{hamburg} by 
utilizing the hyperfine structure of ${}^{87}$Rb.  
Moreover, ensembles consisting of a small 
number of atoms ($\sim 500$) have been realized in~\cite{Schmied}. 
Focusing on ${}^{87}$Rb atoms our dimensionless parameters can 
be expressed in dimensional form by assuming a transversal trapping frequency $\omega_{\perp}=2\pi\times200$ Hz.
Then, all time scales should be rescaled by $4.09$ s and all length scales by $54.7$ $\mu$m. 
This yields an axial trapping frequency $\Omega\approx2\pi$Hz giving an aspect ratio 
$\epsilon=\frac{\Omega}{\omega_{\perp}}=5\times10^{-3}$. These parameters lie within the 
range of applicability of the 1D GP theory according to the criterion \cite{stringari} 
$N\frac{\alpha_{\perp}^4}{\alpha^2 \alpha_{z}^2}=122\gg1$, where $\alpha_{\perp}$, $\alpha_z$ 
denote the oscillator length in the transversal and the axial direction respectively, while $\alpha$ is the 3D $s$-wave 
scattering length. 
As a conclusion due to the time scales and Fig. \ref{Fig:4} showing the suppression of the decay time with increasing number 
of atoms, the aforementioned effects might be observed via time of flight imaging in contemporary experiments that report a 
BEC lifetime of the order of tens of seconds \cite{hamburg}.   

These findings suggest a number of interesting possibilities
for future studies. On the one hand, even at the level
of a single species, identifying states involving
multiple orbitals such as some of the case examples
presented here (e.g., the DWs) would
be of particular interest. On the other hand, extending considerations
to other contexts, involving, e.g., a higher number of
species, or a higher-dimensional setting would
hold considerable promise.
In particular, in the former spinor setting, there
have been argued to exist dark-dark-bright and
dark-bright-bright solitary waves~\cite{hector}, in which
it would be useful to explore not only the spin-independent
collisional features (as is the case here), but also
the {\it spin-dependent} ones.
On the other hand, in the higher dimensional
setting, the role of dark solitons is played by vortices~\cite{kody,pola}.
Such ``vortex-bright'' solitons are on the one hand rather
robust at the mean-field level~\cite{kody}, while also
being topologically protected against splitting
(of the type observed here)~\cite{pola}. Hence, it would
be interesting and relevant to compare/contrast the dynamical
scenarios in the latter case in connection with the former
one, and also in light of recent investigations of vorticity
features in repulsive condensates beyond the mean-field
limit~\cite{marios1,marios2}. Some of these directions are
currently in progress and will be reported in future studies.

\appendix

\section{Initial state preparation} \label{sec:numerics}

To initialize the DB soliton dynamics we embed the mean-field solution to the ML-MCTDHB
ansatz [see Eqs.~(\ref{Eq:5}), (\ref{Eq:6})]. 
The number of particles for each species $N_D$, $N_B$, 
the initial positions $x_j^0$, the velocities $u_j$ and the relative amplitudes $\frac{\eta_j}{\eta_1}$
for the DB solitons are kept fixed to obtain the corresponding mean-field state.
We initialize the algorithm assuming ansatz values for the background chemical potential $\mu^{(0)}$ and inverse widths
$d^{(0)}_j$ of the solitons. The structure of the algorithm is as follows:
 (a) we obtain the mean-field
solution for the background density of the GP equation for $\mu^{(0)}$ and $\mu^{(0)} +\delta \mu$
using the Newton-Krylov algorithm 
(b) we calculate $N_D(\mu)=\int dx|\phi_{\mu}^{D}(x)|^2$,
approximate $\frac{dN_D}{d\mu}$ and update the chemical potential 
according to the Newton-Raphson method   
(c) we iterate (a)-(b) until the number of particles
converges to $N_D$, thus obtaining a new value for the chemical potential $\mu^{(1)}$ (d) we optimize the amplitude of the first 
soliton $\eta_1$ utilizing the Newton-Raphson algorithm (the remaining amplitudes $\eta_{j \neq1}$ and
the inverse widths $d_j$ are fixed due to the specification of the relative amplitudes $\frac{\eta_j}{\eta_1}$ and Eq. 
(\ref{Eq:3})), such that the number of particles for the bright soliton converges to $N_B$, obtaining new values for 
$d_j^{(1)}$. (e) Repeat 
the above process (a)-(d)
by initializing with the newly obtained values of the chemical potential $\mu^{(k)}$ and the inverse widths $d^{(k)}_j$ until 
the convergence criterion
\begin{equation}
\frac{1}{N_S} \sum_{j=1}^{N_S} \frac{\left| D^{(k)}_j-D^{(k-1)}_j \right|}{D^{(k-1)}_j} < 10^{-15},
\end{equation}
is reached.

The solutions obtained by the above process are properly normalized and embedded as the first SPF
for each species $D$ and $B$. The remaining SPFs are initialized in a randomly-generated parity-even
(for the case of collisions) state
and are orthonormalized according to the Gram-Schmidt algorithm. The first species function for
each species is initialized in the state where all the particles reside in the corresponding first
single particle function. The remaining species functions allow for excitations in the randomly generated orbitals.
Finally, the total wavefunction is initialized to the state where only the first species functions for each species is
occupied, i.e. $\lambda_1=1$, and $\lambda_{i\neq1}=0$.
\begin{figure}[t]
\includegraphics[width=0.48\textwidth]{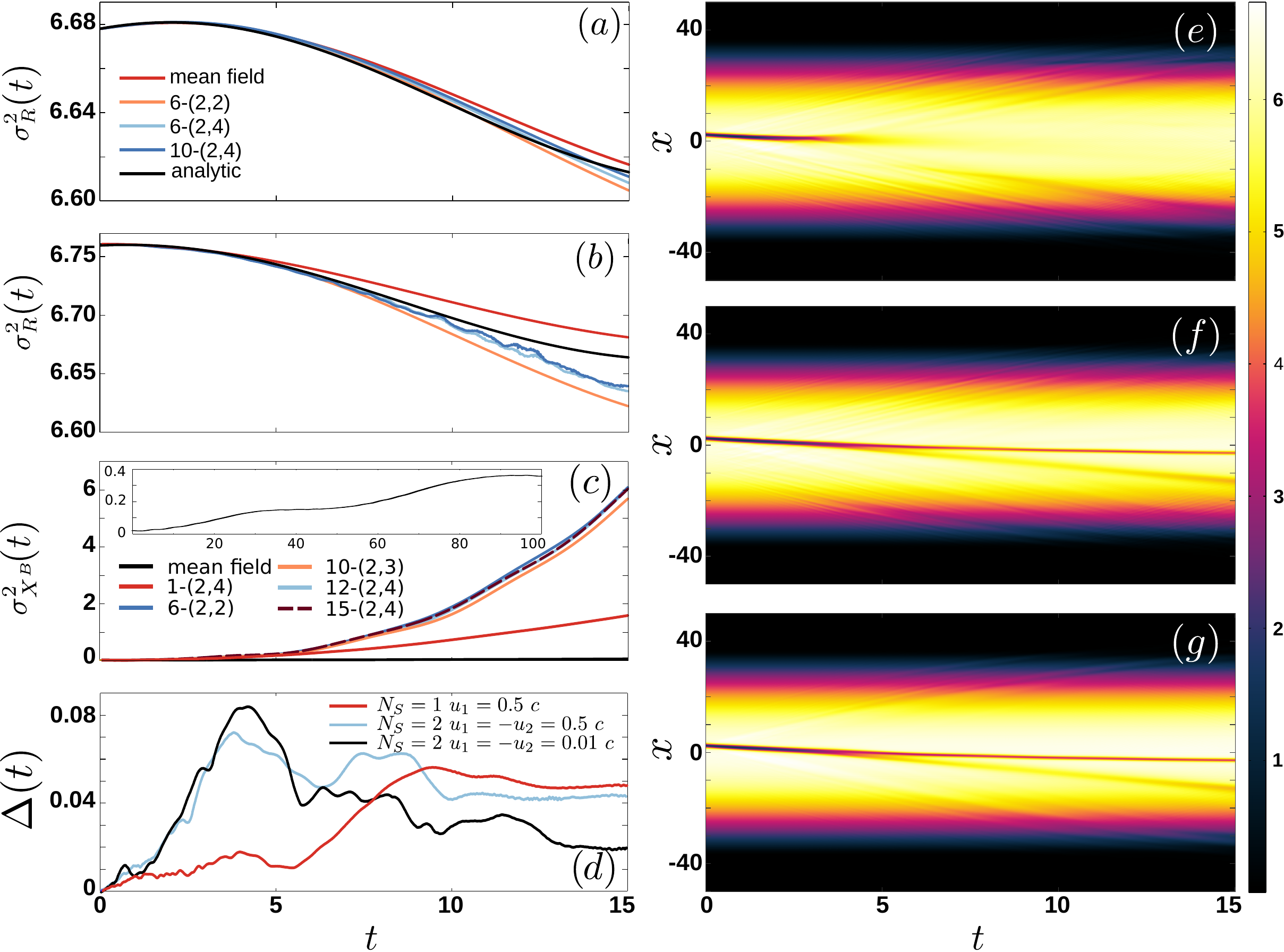}
\caption{(Color online) Evolution of position variance, for a collision between two DB solitary waves, using different correlated 
approximations, the mean-field ansatz 
and the analytical result (see legend) in the case of ($a$) fast velocities with $u_1/c=0.5$ and ($b$) slow velocities with 
$u_1/c=0.01$. 
($c$) Time evolution of the position variance of the center of mass for the bright component $\sigma_{X_B}^2(t)$ calculated for different 
correlated approximations, the species mean-field 1-(2,4) and the mean-field case (see legend). In the inset we show solely $\sigma_{X^B}^2(t)$ within the mean-field 
approximation. 
($d$) Evolution of the relative error $\Delta(t)$ (see main text) of the one-body density of the dark component at the 
``core'' of the DB solitary wave for the dynamics of a single and two DB structures (see legend). ($e$), ($f$), ($g$) One-body density evolution for the fast oscillation ($u_1/c=0.5$) of 
a DB solitary wave within the 1-(3,4), 10-(3,4)  
and 10-(2,4) approximations respectively.  
The remaining parameter values are the same as in Fig.~\ref{Fig:5} and Fig.~\ref{Fig:8} respectively.} \label{Fig:11}
\end{figure} 

\section{Remarks on Convergence} \label{sec:numerics1}

In the present Appendix we briefly discuss the main features of our computational method (ML-MCTDHB) and then demonstrate the 
convergence of our results. 

Within ML-MCTDHB the total many-body wave function is expanded with respect to a time-dependent variationally optimized 
many-body basis.  
The latter allows us to obtain converged results with a reduced number of basis states compared to expansions relying on a 
time-independent basis. 
Also, the symmetry of the bosonic species is explicitly employed. 
Finally, the multi-layer ansatz for the total wave function is based on a coarse-graining cascade, where 
strongly correlated degrees of freedom are grouped together and treated as subsystems, which mutually couple to each other. 
The latter enables us to adapt the ansatz to intra- and inter-species correlations and makes ML-MCTDHB a highly
flexible tool for simulating the dynamics of e.g. bipartite systems. 

The primitive underlying basis for both the Newton-Krylov algorithm and the ML-MCTDHB simulations corresponds to a sine 
discrete variable representation (DVR) with 1200
grid points. 
The number of species functions is always kept equal to $M_D=M_B=M$. To obtain the corresponding order of approximation we use 
a specific 
number of species functions $M$ and single
particle functions $m_D=m_B$ (given within the main text). 
To track the numerical error and guarantee the accurate performance of 
the numerical integration for the ML-MCTDHB equations of motion we impose the following overlap criteria $|\langle \Psi |\Psi \rangle -1| < 10^{-10}$ and
$|\langle \varphi_i |\varphi_j \rangle -\delta_{ij}| < 10^{-10}$ for the total wavefunction and the SPFs respectively. 
For the same reasons, also, for the two DB soliton dynamics we ensured that $|\braket{\Psi_{MB}|\hat{\mathcal{P}}|\Psi_{MB}}-1|<10^{-6}$, where 
$\hat{\mathcal{P}}$ denotes the many-body parity operator, i.e.  
$\hat{\mathcal{P}}\Psi_{MB}(\vec{x}^{D},\vec{x}^{B})=\Psi_{MB}(-\vec{x}^{D},-\vec{x}^{B})$. 

Next, let us elaborate in more detail on the convergence of our simulations. 
To demonstrate the level of our many-body wavefunction truncation scheme we first show the behaviour of the center-of-mass variance calculated both analytically (see below)  
and numerically. 
Then, we present the relative error at the core of the dark component (i.e. minimizing the effect of the background) between the 10-(2,4) and 10-(3,4) approximations i.e. by adding 
one more SPF in the dark component. 
We remark that the harmonic oscillator potential allows for the separation of the center of mass (CM), $R=\frac{1}
{N}\sum_ix_i$, from the relative coordinates $r_i=x_{i+1}-x_{i}$.   
This separation enables to reduce the $N$-body ($N=N_D+N_B$) interacting problem to an interacting $N-1$-body problem in the relative 
coordinates, and a non-interacting one for the CM coordinate.
Then the many-body Hamiltonian reads 
\begin{equation}
\hat H = H_R +\sum_{i=1}^{N_D+N_B-1} H_{r_i},
\end{equation} 
where $H_R=-\frac{1}{2 N} \frac{d^2}{d R^2}+\frac{1}{2} N m \omega^2 R^2$ is the single particle harmonic oscillator 
Hamiltonian of the center of 
mass, and $H_{r_i}$ is the interaction Hamiltonian in the relative coordinates.  
However, our calculations within ML-MCTDHB have been performed in the lab frame and as a consequence do not utilize the 
aforementioned separation of variables. 
Note that the ML-MCTDHB ansatz (as well as the corresponding mean-field ansatz) does not trivially respect the separation 
between the CM and relative frame. 
However, as we shall show below our results can capture the decoupling of the CM motion for the entire $N_D+N_B$ bosonic cloud.  
To judge about relative deviations of the ML-MCTDHB propagation with the full Schr{\"o}dinger equation (and consequently about 
convergence) we compare the ML-MCTDHB obtained 
evolution of the center of mass coordinate to the analytical one. 
Since we are interested only in the one and two-body densities we are able to calculate the first and second statistical 
moments of the center of mass.  
The second moment of the CM position (position variance) reads 
\begin{equation} 
\begin{split}
&\sigma_R^2(t)=\frac{N_D \langle{x_D^2}\rangle(t) + N_B \langle{x_B^2}\rangle(t)}{(N_D+N_B)^2}
\\&+ \frac{N_D (N_D-1)}{(N_D+N_B)^2} \langle{x_D x'_D}\rangle(t)+\frac{2 N_D N_B}{(N_D+N_B)^2} \langle{x_D x'_B}\rangle(t)
\\&+\frac{N_B (N_B-1)}{(N_D+N_B)^2} \langle{x_B x'_B}\rangle(t)
-\left( \frac{N_D \langle{x_D}\rangle(t) + N_B \langle{x_B}\rangle(t)}{N_D+N_B} \right)^2, 
\end{split}  
\end{equation} 
where $\langle{x_i}\rangle$ denotes the mean 
position of the $i$-th species
and $\langle{x_ix'_j}\rangle$ refers to the two-body spatial moment of the position of particles of species $i$ and $j$ 
respectively~\cite{klaiman1,klaiman2}.  

By using the Ehrenfest theorem on the center of mass Hamiltonian $H_R$ we obtain the exact evolution of the CM position 
variance 
\begin{equation}
\begin{split}
 \sigma_R^2(t)=&\left[\langle{R^2}\rangle(0)-\langle{R}^2\rangle(0)\right] \cos^2 \omega t \\ %\nonumber 
&+ \frac{1}{\omega^2} \left[\langle{P^2}\rangle(0)-\langle{P}^2\rangle(0)\right] \sin^2 \omega t \\ %\nonumber
&+\frac{1}{2 \omega}\langle{ R P + P R }\rangle(0) \sin 2 \omega t \\ %\nonumber 
&-\frac{1}{\omega}\langle{ R}\rangle(0)\langle{ P}\rangle(0) \sin 2 \omega t. \label{Eq:B4} 
\end{split} 
\end{equation} 
$R$, $P$ denote the spatial coordinate and the momentum operators 
acting on the CM degree of freedom. 
This latter expression offers the opportunity to directly measure the deviation between the correlated approach, the
mean-field ansatz, and the analytical result \cite{Cosme}. 
To expose this deviation we numerically 
calculate the position variance (using different number of orbitals or species functions) and compare with Eq.~(\ref{Eq:B4}). 
This comparison is shown, for the case of a collision between two 
DBs, in Fig.~\ref{Fig:11} ($a$), ($b$) for fast ($u/c=0.5$) and slow ($u/c=0.01$) velocities respectively. 
As it can be seen all the correlated obtained results follow the behaviour of the analytical result and therefore can be 
considered trustworthy.  
The observed deviation at long propagation times is of the order of $0.1\%$ lying within the numerical error of the lattice discretization $0.6\%$. 
Note that increasing the number of SPFs the correlated approach (see e.g. the 15-(2,4) approximation) 
tends to a better agreement with 
the analytical result. 
As expected, the mean-field result (when compared to the correlated approximation) possesses a larger deviation with the 
analytics.  
Also the case of slow velocities shows a comparatively larger deviation (than the case of fast velocities) as it possesses a 
larger amount of depletion. 
Next, we illustrate the convergence of our numerical results for another two-body observable, namely the 
position variance for the 
%center of mass 
CM of the bright soliton 
\begin{equation}
\begin{split}
 \sigma_{X^B}^2(t)=&\braket{\Psi_{MB}(t)|(X^B)^2|\Psi_{MB}(t)}\\&-\braket{\Psi_{MB}(t)|X^B|\Psi_{MB}(t)}^2, \label{Eq:B5} 
\end{split} 
\end{equation} 
where $X^B=\left(1/N_B\right)\left(x_1^B+...+x_{N_B}^B\right)$ and $(X^B)^2=\left(1/N_B\right)\sum_i^{N_B}(x_i^B)^2+\left(2/N_B\right)\sum_{i<j}x_i^Bx_j^B$. 
Note here that due to the coupling of the CM of the $B$ component with the entirety of the $D$ component no 
analytic form can be extracted for $\sigma_{X^B}^2(t)$. 
Fig. \ref{Fig:11} ($c$) presents $\sigma_{X^B}^2(t)$ for an increasing number of species functions and/or  
number of SPFs of both the dark and bright component.  
As it can be seen the CM for the bright soliton expands during the evolution both within the mean-field (see also the inset)  
and within the correlated approximation. 
However, in the correlated case the expansion possesses a larger amplitude when compared to the mean-field approximation.   
On the convergence side, as shown for increasing number of entanglement modes $M$ the corresponding $\sigma_{X^B}^2(t)$ graphs are 
almost indistinguishable  
(see brown dashed and light blue lines respectively). It is important to note here that the species mean-field case 1-(2,4) 
differs significantly from the case 15-(2,4). The latter signals the necessity for the inclusion 
of entanglement modes $M$ in order to fairly capture the DB soliton dynamics (see also below). 
At the same time, the near-coincidence of the different graphs with the number
of modes used is a strong indication of the convergence and the
trustworthiness of our results. 

Note that within the current calculations we do not ensure the convergence of the ML-MCTDHB  
simulations at the level of the ($N_D+N_B$)-body wavefunction, as such investigations are computationally prohibitive 
for the large particle numbers considered here.  
However, we showcase the robustness of the emerging structures in the 
beyond mean-field dynamics for lower order properties e.g. the one-body density matrix and related one-body quantities.  
To demonstrate the order of approximation of our results we present the evolution of the relative error integrated within the ``core'' of the 
dark soliton. In particular, we calculate the relative one-body density difference of the dark component  
$\Delta(t)=\frac{1}{L} \int_R dx \frac{\abs{\rho^{(1),D}_{10-(2,4)}(x;t)-\rho^{(1),D}_{10-(3,4)}(x;t)}}{\rho^{(1),D}_{10-(3,4)}(x;t)}$,  
where the domain $R$ is such that $x \in [-10,10]$ and refers to the ``core'' of the dark component; $L=20$ is the length of the integrated domain $R$.  
More precisely, here, we compare the approximations 10-(2,4) and 10-(3,4), i.e. we include one more SPF in the dark components, thus enlarging the available Hilbert space ($\mathcal{H}^{DB}$) for 
the simulation by taking into account 462370 coefficients (in contrast to the $10800$ used in the $10-(2,4)$ case).    
Fig. \ref{Fig:11} ($c$) shows $\Delta(t)$ for the oscillation of a DB solitary wave and the cases of both slow and fast collisions between two DBs respectively.
As can be seen, for the oscillation of a DB solitary wave $\Delta(t)$ before the decay (i.e. $t<t_D$) lies below $2\%$, while for $t>t_D$ it increases in a linear manner up to $6\%$.  
On the other hand, for the case of a slow collision between two DBs $\Delta(t)$ increases during the evolution 
reaching a maximum value of the order of $8.5\%$ during the decay process, while after the decay $\Delta(t)$ decreases to $5\%$. 
For the fast soliton collision the respective error reaches a maximum value of about $5\%$. 
Finally, in Figs. \ref{Fig:11} ($d$), ($e$), ($f$) we present the one-body density evolution of the dark component for the case of a fast oscillation of a DB  
solitary wave for different approximations, namely 1-(3,4), 10-(3,4) and 10-(2,4) respectively. It is observed that in the species mean-field case, i.e. 1-(3,4) approximation, 
the two DB structures after the decay produce a jet-like structure and all DB entities are lost.  
The latter indicates the inescapable necessity of taking more than one species functions for the described excitation dynamics, 
due to the emergence of the inter-species modes of entanglement.  
The convergence of our results for a given number of SPFs ($m_D,m_B)$ and increasing number 
of $M$ is observed, verifying that the nature of excitations does not change for increasing number of available inter-species 
modes of entanglement $M$. 
On the other hand, by inspecting the dynamics within the approximations 10-(2,4) and 10-(3,4) we observe a similar qualitative behaviour (confirming also the above described relative 
error [see Fig. \ref{Fig:11} ($c$)]) before and after the decay process. 
Similar investigations have also been performed with respect to the bright component SPFs ($m_B$) not included here for brevity reasons. 

Finally, we remark that most of the presented results have been performed within the 15-(2,4) approximation (i.e. using 12780 coefficients), 
which according to the aforementioned criteria is an adequate 
approximation for the description of the induced dynamics at the one-body density level. 
Summarizing, we have observed the same emerging structures during the dynamics for all of the presented orbital/species 
functions combinations except for the species mean field case (namely the 1-(3,4) approximation).

\section*{Acknowledgements} 
The authors would like to thank S. Kr{\"o}nke for fruitful discussions. 
The authors G.M.K, S.I.M. and P.S. gratefully acknowledge financial support by the Deutsche Forschungsgemeinschaft 
(DFG) in the framework of the
SFB 925 ``Light induced dynamics and control of correlated quantum
systems'' and the excellence cluster ``The Hamburg Center for Ultrafast Imaging: Structure, Dynamics and Control
of Matter at the Atomic Scale''. G.C.K and P.S. gratefully acknowledge financial support by DFG in the framework of the grant 
SCHM 885/26-1. P.G.K. gratefully acknowledges the
support of NSF-DMS-1312856, NSF-PHY-1602994, the
Alexander von Humboldt Foundation, and the ERC under
FP7, Marie Curie Actions, People, International Research
Staff Exchange Scheme (IRSES-605096).

{}


\begin{thebibliography}{60}
\bibitem{note_ent} Two widely used measures in the literature to quantify bipartite entanglement are the von-Neumann 
entropy and the concurrence. The von-Neumann entropy is expressed 
by $S(\Psi_{MB}(t))=-Tr(\rho^{\sigma}log(\rho^{\sigma}))=-\sum_k\lambda_k(t)\log(\lambda_k(t))$, while the concurrence is $D(\Psi_{MB}(t))=(Tr\sqrt{\rho^{\sigma}})^2-1=2 
\sum_{i<j}\sqrt{\lambda_i(t) \lambda_j(t)}$,  
where $\rho^{\sigma}(t)=\sum_k\lambda_k(t)\ket{\Psi_k^{\sigma}(t)}\bra{\Psi_k^{\sigma}(t)}$ is the $N$-body density matrix and 
$\Psi_k^{\sigma}(t)$ denotes the $k$-th species function of the $\sigma$ species. Both of the above-mentioned quantities vanish only in 
the factorized case, i.e. non entangled case.
\bibitem{note1}
In the cited papers \cite{Sven,Lushuai} the more general ML-MCTDHB ansatz of a mixture composed of an arbitrary number of 
bosonic components has been used.
The form of the ML-MCTDHB ansatz presented in this study 
utilizes the Schmidt decomposition that holds for binary mixtures~\cite{Johannes}.
\bibitem{note_vel} Note here that, in the general case, $v_{init}\neq c \sin \alpha_1$ due to initial sound wave emissions.
\end{thebibliography}
\end{document}